\documentclass[aps,pra,twocolumn,floatfix,superscriptaddress,nofootinbib]{revtex4-2}
\usepackage[T1]{fontenc}
\usepackage{amsmath}
\usepackage{amsbsy}
\usepackage{amssymb}
\usepackage{dsfont}
\usepackage[utf8]{inputenc}
\usepackage[english]{babel}
\usepackage[pdftex,dvipsnames,usenames]{xcolor}
\usepackage[colorlinks=true,urlcolor=blue,citecolor=blue,linkcolor=blue]{hyperref}
\usepackage{lipsum}
\usepackage{graphicx, color}

\newcommand{\ket}[1]{\left|#1\right\rangle}

\DeclareMathOperator{\BesselM}{I}
\DeclareMathOperator{\erfc}{erfc}
\DeclareMathOperator{\corr}{corr}

\begin{document}
\title{Circular-beam approximation for quantum channels in a turbulent atmosphere}

\author{I. Pechonkin}
\affiliation{Quantum Optics and Quantum Information Group, Bogolyubov Institute for Theoretical Physics of the National Academy of Sciences of Ukraine, Vulytsia Metrolohichna 14b, 03143 Kyiv, Ukraine}

\author{M. Klen}
\affiliation{Quantum Optics and Quantum Information Group, Bogolyubov Institute for Theoretical Physics of the National Academy of Sciences of Ukraine, Vulytsia Metrolohichna 14b, 03143 Kyiv, Ukraine}

\author{A. A. Semenov}
\affiliation{Quantum Optics and Quantum Information Group, Bogolyubov Institute for Theoretical Physics of the National Academy of Sciences of Ukraine, Vulytsia Metrolohichna 14b, 03143 Kyiv, Ukraine}
\affiliation{Department of Theoretical and Mathematical Physics, Kyiv Academic University, Boulevard Vernadskogo  36, 03142  Kyiv, Ukraine}
\affiliation{Department of Mathematics, Kyiv School of Economics, Vulytsia Mykoly Shpaka 3, 03113  Kyiv, Ukraine}

\begin{abstract}
The evolution of quantum states of light in free-space channels is strongly influenced by atmospheric turbulence, posing a significant challenge for quantum communication.
The transmittance in such channels randomly fluctuates.
This effect is commonly described by the probability distribution of transmittance (PDT). 
The elliptic-beam approximation provides an analytical model for the PDT, showing good agreement with experimental and simulation data within a specific range of channel parameters. 
In this work, we introduce the circular-beam approximation---a simplified alternative that offers satisfactory accuracy while significantly reducing computational complexity. 
Our method naturally leads to a technique for determining the model parameters from the first two moments of the transmittance.
This approach eliminates the model misspecification bias inherent in the elliptic-beam approximation and significantly extends the applicability range of the PDT model, providing a practical tool for characterizing atmospheric channels in quantum applications.
\end{abstract}

\maketitle


\section{Introduction}

The distribution of quantum light through free-space channels attracted considerable attention due to its relevance for applications such as quantum secure communication~\cite{gisin02,Xu2020,Pirandola2020,Renner2023}, supported by numerous successful experiments; see, e.g., Refs.~\cite{ursin07,elser09,heim10,fedrizzi09,capraro12,yin12,ma12,peuntinger14,vasylyev17,Jin2019,nauerth13,wang13,bourgoin15,vallone15,dequal16,vallone16,carrasco-casado16,takenaka17,liao17,yin17,gunthner17,ren17,yin17b,Liao2018,Vasylyev2019,Ecker2021,Ecker2023}. 
In a typical scenario, quantum states of light are transmitted from one station (the transmitter) to another (the receiver). 
However, this task cannot be performed perfectly due to environmental noise, including stray light and atmospheric turbulence. 
The latter causes random fluctuations in the refractive index, which, in turn, randomly distort the temporal and spatial profiles of light modes.

To understand how this effect influences the quantum state of light, it is first necessary to specify the method of quantum information encoding at the transmitter and its detection at the receiver. 
In this context, several encoding techniques can be considered, including encoding via spatial~\cite{paterson05,pors11,Roux2013,DAmbrosio2012,vallone2014,Roux2015,Leonhard2015,krenn2016,Lavery2018,Sorelli2019,Klug2023,Bachmann2023} and temporal~\cite{Klen2024} mode structures, polarization modes~\cite{ursin07,fedrizzi09,semenov10,gumberidze16}, as well as states of a quasimonochromatic mode in the form of a Gaussian light beam with subsequent photocounting measurements~\cite{perina73,milonni04} or homodyne detection~\cite{usenko12,guo2017,papanastasiou2018,derkach2020b,hosseinidehaj2019,ShiyuWang2018,chai2019,Derkach2021,pirandola2021,hosseinidehaj2021,pirandola2021b,pirandola2021c,peuntinger14,Wang2018,hofmann2019,zhang2017,villasenor2021,bohmann16,hosseinidehaj15a,elser09,heim10,usenko12,peuntinger14,heim14}, among others. 
The effect of atmospheric turbulence on the quantum state in the latter two encoding schemes can be characterized by random fluctuations in the transmittance efficiency: as the beam shape randomly fluctuates at the receiver, the fraction of light passing through the receiver aperture also varies accordingly. 
In the following, we focus our analysis on this scenario.

Let the quantum states at the receiver and transmitter stations be described by the Glauber–Sudarshan $P$ functions \cite{glauber63c,sudarshan63} $P_{\text{out}}\left(\alpha\right)$ and $P_{\text{in}}\left(\alpha\right)$, respectively.
As discussed in Ref.~\cite{semenov09} (see also a detailed derivation in Ref.~\cite{Klen2024}), these states are related to each other as
	\begin{align}\label{Eq:InOut}
		P_{\text{out}}\left(\alpha\right)=\int_{0}^1 d\eta \mathcal{P}(\eta) \frac{1}{\eta} P_{\text{in}}\left( \frac{\alpha}{\sqrt{\eta}} \right).
	\end{align}
Here, $\eta\in[0,1]$ is a random transmittance and $\mathcal{P}(\eta)$ is the probability distribution of transmittance (PDT)---a function characterizing quantum free-space channels.
Based on this input–output relation, the transmission of quantum properties through atmosphere channels \cite{semenov12,vasylyev12,semenov10,gumberidze16,bohmann16,hosseinidehaj15a,bohmann17a,hosseinidehaj15b} and quantum communication protocols \cite{Wang2018,vasylyev18,usenko12,guo2017,papanastasiou2018, derkach2020b,hosseinidehaj2019,ShiyuWang2018,chai2019, Derkach2021,pirandola2021,hosseinidehaj2021,pirandola2021b,pirandola2021c,hofmann2019,zhang2017,villasenor2021} has been analyzed.

These results demonstrate that having an appropriate PDT model is crucial for the accurate description of free-space quantum channels.
The simplest analytical model is given by the log-normal distribution \cite{perina73,milonni04}, which must be properly truncated to the domain $\eta\in[0,1]$.
Another heuristic model, which shows a good agreement with numerical simulations, is based on the beta distribution; see Ref.~\cite{klen2023}.
Assuming statistically independent contributions of beam–wandering and beam-spot distorsion, a model based on the law of total probability was derived in Ref.~\cite{vasylyev18}. 
The beam-wandering model, assuming a Gaussian beam profile at the receiver and considering the effect of beam wandering, was introduced in Ref.~\cite{vasylyev12}.
Its modification, which additionally accounts for random fluctuations of an elliptically shaped beam, was presented in Ref.~\cite{vasylyev16}. 
The latter provides a good fit to experimental data and remains effective under diverse weather conditions~\cite{vasylyev17}.

Nevertheless, two issues are associated with the elliptic-beam model: (i) it requires complex calculations involving approximations and numerical simulations; (ii) although it provides a good fit to experimental data over a broad range of channel parameters, numerical simulations in Ref.~\cite{klen2023} have shown that the employed parameter evaluation method introduces a significant bias in the distribution mode, causing the model to perform reliably only within a narrow range of channel parameters.
The aim of this paper is to address these issues. 
We propose a simplified version of the elliptic-beam approximation, assuming that the beam retains a Gaussian profile but with a circular shape of a random radius---the circular-beam approximation. 
While this model reduces computational complexity, it still exhibits a notable bias in the distribution mode across a significant range of parameter values. 
This limitation is primarily due to the strongly non-Gaussian character of the beam shape at the receiver. 
To overcome this model–misspecification bias, we introduce an alternative parameter evaluation method based on matching the first two moments of the transmittance distribution.

The rest of the paper is organized as follows. 
In Sec.~\ref{Sec:Preliminaries} we provide the preliminary information necessary for introducing our method. 
Section~\ref{Sec:Circbeam} presents the circular-beam model and the associated method for determining the PDT parameters, which eliminates model misspecification bias.
Analytical expressions for the model parameters, derived within the framework of the Huygens--Kirchhoff phase approximation, are given in Sec.~\ref{Sec:Analytical}. 
Examples of the PDT and validation of our models through numerical simulations are discussed in Sec.~\ref{Sec:Examples}. 
In Sec.~\ref{Sec:Application} we present examples demonstrating the application of our techniques to describe the transmission of nonclassical properties of quantum light through free-space channels. 
Finally, a summary and conclusions are provided in Sec.~\ref{Sec:Conclusions}.


\section{Preliminaries}
\label{Sec:Preliminaries}

In this section we provide the reader with a preliminary background on free-space channels~\cite{Tatarskii,Tatarskii2016,Fante1975,Fante1980,Andrews_book}, which is essential for formulating our approach. 
We also summarize key aspects of the beam-wandering model~\cite{vasylyev12}, which plays a crucial role in our technique.
Our basic assumption is that quantum states of light are prepared in a quasi-monochromatic mode, representing a pulsed Gaussian beam.
Since Eq.~\eqref{Eq:InOut} holds for both quantum and classical fields, the derivation of the PDT can be carried out by analyzing the transmittance of a classical quasi-monochromatic mode through the receiver aperture; for details, see Appendix~A of Ref.~\cite{klen2023}.
 
Let $u(\textbf{r};z)$ be the beam amplitude, where $\textbf{r}=(x,y)$ and $z$ are the transverse and axial coordinates in the propagation direction, respectively.   
In the paraxial approximation, this amplitude satisfies the equation (see, e.g., Ref.~\cite{Fante1975})
    \begin{align}\label{Eq:paraxial}
		2ik\frac{\partial u(\textbf{r};z)}{\partial z}+ \Delta_{\textbf{r}} u(\textbf{r};z)+2k^2 \delta n(\textbf{r};z)u(\textbf{r};z)=0.
	\end{align}
Here $k$ is the wave number, $\Delta_{\textbf{r}}$ is the transverse Laplace operator, and $\delta n(\textbf{r};z)$ represents small random fluctuations in the index of refraction due to atmospheric turbulence.
The Gaussian beam is defined by the boundary conditions
	\begin{align}\label{Eq:BoundaryConditions}
		u(\mathbf{r};0)=\sqrt{\frac{2}{\pi
		W_0^2}}\exp\Bigl[-\frac{\mathbf{r}^2}{W_0^2}{-}\frac{ik}{2F_0}\mathbf{r}^2\Bigr],
	\end{align}
where $W_0$ and $F_0$ are the beam-spot radius and the wave–front radius at the transmitter, respectively.

The second-order correlation function of $\delta n(\textbf{r};z)$ in the Markovian approximation is defined as 
	\begin{align}\label{Eq:Correlator_n_n}
		\langle\delta n&(\mathbf{r}_1;z_1)\delta n(\mathbf{r}_2;z_2)\rangle\nonumber\\
		&=\int_{\mathbb{R}^2}d^2\boldsymbol{\kappa}\Phi_n(\boldsymbol{\kappa};\kappa_z{=}0)e^{i\boldsymbol{\kappa}\cdot(\mathbf{r}_1-\mathbf{r}_2)}\delta(z_1-z_2),
	\end{align}
where $\Phi_{n}(\boldsymbol{\kappa};\kappa_z)$ is the turbulence power spectral density.
A commonly used model is the modified von K\'arm\'an--Tatarskii spectrum, given by
	\begin{align}\label{Eq:Karman}
		\Phi_n(\boldsymbol{\kappa};\kappa_z)=\frac{0.033 C_n^2\exp\left[-\left(\frac{\kappa\ell_0}{2\pi}\right)^2\right]}{\left(\kappa^2+L_0^{-2}\right)^{11/6}}.
	\end{align}
Here $C_n^2$ is the index-of-refraction structure constant characterizing the local turbulence strength, $L_0$ and $\ell_0$ are outer and inner scales of turbulence, respectively, and $\kappa=\sqrt{\boldsymbol{\kappa}^2+\kappa_z^2}$.
For analytical approximations, we set $L_0=+\infty$ and $\ell_0=0$, which yields the Kolmogorov spectrum.

The intensity of the light at the receiver station is given by
    \begin{align}\label{Eq:Intensity}
    	I(\textbf{r};L)=|u(\textbf{r};L)|^2,
    \end{align}
where $z=L$ is the axial coordinate of the aperture plane (channel length).   
Random fluctuations in $\delta n(\textbf{r};L)$ lead to random fluctuations in $I(\textbf{r};L)$.
For the purposes of our paper, it is sufficient to consider only three random variables that characterize this random field.
The first variable is the random transmittance efficiency, defined as
    \begin{align}\label{Eq:eta}
		\eta =\int_{\mathcal{A}} d^2\textbf{r} I(\textbf{r};L),
	\end{align}
where $\mathcal{A}$ denotes the aperture opening.
The second variable is the beam-centroid position $\mathbf{r}_0=(x_0,y_0)$ given by
    \begin{align}\label{Eq:centroidCoord}
		\textbf{r}_0 = \int_{\mathds{R}^2} d^2\textbf{r}\, \textbf{r}\, I(\textbf{r};L) .
	\end{align}
The third variable, 
	\begin{align}\label{Eq:simpleS}
		S  = 4\int_{\mathds{R}^2} d^2\textbf{r} (x-x_0)^2 I(\textbf{r};L),
	\end{align}
is the squared instant beam-spot radius.

We are interested in the first two moments of these random variables.
These moments can be conveniently expressed in terms of the second- and fourth-order field correlation functions, defined as 
    \begin{align}\label{Eq:G2}
		\Gamma_2(\textbf{r};L)= \left\langle I(\textbf{r};L) \right\rangle
	\end{align}
and	
	\begin{align}\label{Eq:G4}
		\Gamma_4(\textbf{r}_1, \textbf{r}_2;L)= \left\langle I(\textbf{r}_1;L)I(\textbf{r}_2;L) \right\rangle,
	\end{align}
respectively.
These functions can be obtained using methods of classical atmospheric optics; see, e.g., discussions in Sec.~\ref{Sec:Analytical}.
The first and second moments of the transmittance are given by
    \begin{align}\label{Eq:etaAvrg}
		\left\langle\eta\right\rangle =\int_{\mathcal{A}} d^2\textbf{r}\, \Gamma_2 \left(\textbf{r}; L\right),
	\end{align}
	\begin{align}\label{Eq:eta2Avrg}
		\left\langle\eta^2\right\rangle =\int_{\mathcal{A}} d^2\textbf{r}_1 d^2\textbf{r}_2\, \Gamma_4 \left(\textbf{r}_1, \textbf{r}_2; L\right).
	\end{align}
We chose the coordinate system such that $\left\langle\mathbf{r}_0\right\rangle=0$.
Then the variance of a beam-centroid coordinate is given by
	\begin{align}\label{Eq:bwVariance}
		\sigma^2_{\text{bw}} &= \left\langle \Delta x_0^2 \right \rangle\\
		&=\left\langle x_0^2 \right\rangle = \int_{\mathds{R}^4} d^2\textbf{r}_1 d^2\textbf{r}_2 x_1 x_2 \Gamma_4(\textbf{r}_1, \textbf{r}_2;L). \nonumber
	\end{align}
The mean–squared beam-spot radius reads	
	\begin{align}\label{Eq:S}
		\left\langle S \right\rangle=4\left(\int_{\mathds{R}^2} d^2\textbf{r} x^2 \Gamma_2(\textbf{r};L) - \left\langle x_0^2 \right\rangle \right).
	\end{align}
The second moment of this random variable requires a more involved analysis, which is presented in Appendix~\ref{App:MatchS}.

In the circular-beam approximation presented in this paper, we assume that the instant intensity has a Gaussian form:
    \begin{align}\label{Eq:IntensityGauss}
		I(\textbf{r},L)= \frac{2}{\pi S} \exp\left( -\frac{2}{S} |\textbf{r}-\textbf{r}_0|^2 \right).
	\end{align}
Following a procedure similar to that used for the beam-wandering model~\cite{vasylyev12} and assuming that the transmittance fluctuations arise solely from centroid fluctuations, we obtain the conditional PDT for a fixed value of $S$,
    \begin{align}\label{Eq:BWPDT}
		\mathcal{P}(\eta|S)=\frac{R^2(S)}{\sigma^2_{\text{bw}} \eta \lambda(S) } \left(\ln\frac{\eta_0(S)}{\eta} \right)^{2/\lambda(S)-1} \\ 
		\times\exp\left[ -\frac{R^2(S)}{2\sigma^2_{\text{bw}}} \left(\ln\frac{\eta_0(S)}{\eta} \right)^{2/\lambda(S)} \right]\nonumber
	\end{align}
for $\eta\in\left[0, \eta_0  \right]$ and $\mathcal{P}(\eta|S)=0$ otherwise. 
Here
	\begin{align}\label{Eq:eta0}
		\eta_0(S)=1-\exp\left(-\frac{a^2}{S}\right)
	\end{align}
is the maximal transmittance for the given $S$,	
	\begin{align}\label{Eq:lambda}
		\lambda(S) &= 8 \frac{a^2}{S} \frac{\exp\left(-4\frac{a^2}{S} \right) \BesselM_1\left(4\frac{a^2}{S} \right)}{1-\exp\left(-4\frac{a^2}{S} \right) \BesselM_0\left(4\frac{a^2}{S} \right)} \\
		&\times \left[\ln\left( \frac{2\eta_0(S)}{1-\exp\left(-4\frac{a^2}{S} \right) \BesselM_0\left(4\frac{a^2}{S} \right)} \right)\right]^{-1}
	\end{align}
is the shape parameter,
	\begin{align}\label{Eq:R}
		R(S)=a \left[\ln\left( \frac{2\eta_0(S)}{1-\exp\left(-4\frac{a^2}{S} \right) \BesselM_0\left(4\frac{a^2}{S} \right)} \right)\right]^{-1/\lambda(S)}
	\end{align}
is the scale parameter, $a$ is the aperture radius, and $\BesselM_n(x)$ is the modified Bessel function. 

By setting $S=\left\langle S \right\rangle$ in Eq.~(\ref{Eq:BWPDT}), we recover the beam-wandering PDT \cite{vasylyev12}, which neglects fluctuations in the beam-spot radius.
A more realistic model, the elliptic-beam approximation~\cite{vasylyev16}, still assumes a Gaussian beam profile but with an elliptical shape.
This approach involves a more complex analysis due to the increased number of fluctuating parameters.
In this work we aim to simplify the description by assuming that the beam retains a circular Gaussian shape while allowing the squared beam-spot radius $S$ to fluctuate.


\section{Circular-beam approximation}
\label{Sec:Circbeam}

In this section we introduce the main concept of the circular-beam approximation and describe a method for determining its parameters by matching the first two moments of the transmittance $\eta$.
On the one hand, this approach provides a straightforward simplification of the elliptic-beam approximation presented in Ref.~\cite{vasylyev16}, thereby avoiding the need for the more involved analytical and numerical calculations required in that model.
On another hand, this simplification enables a direct matching of the model parameters to the transmittance moments, thus reducing the model misspecification bias that arises in the elliptic-beam approximation over a wide range of aperture radii $a$, see Ref.~\cite{klen2023}.

We start with Eq.~(\ref{Eq:BWPDT}) for the conditional PDT with a given value of the squared beam-spot radius $S\in[0,+\infty)$.
Assuming that this random variable follows a probability distribution $P(S)$, the total PDT is given by
    \begin{align}\label{Eq:fullprob}
		\mathcal{P}(\eta)=\int_{0}^\infty dS \mathcal{P}(\eta|S) P(S).
	\end{align} 
As in the elliptic-beam model, we assume for the circular-beam model that $S$ follows a log-normal distribution  \cite{crow1988}:
     \begin{align}\label{Eq:lognorm}
		P(S)=\frac{1}{S\sigma\sqrt{2\pi}}\exp{-\frac{\left(\ln{S}-\mu\right)^2}{2\sigma^2}}.
	\end{align} 
Here $\mu$ and $\sigma^2$ are the parameters of the log-normal distribution.
Thus, the circular-beam PDT is fully characterized by three parameters: the variance of the beam-centroid coordinate $\sigma^2_{\text{bw}}$ [cf. Eq.~(\ref{Eq:bwVariance})] and the log-normal parameters $\mu$ and $\sigma^2$.  

\begin{figure}[ht!]
		\centering
		\includegraphics[width=1\linewidth]{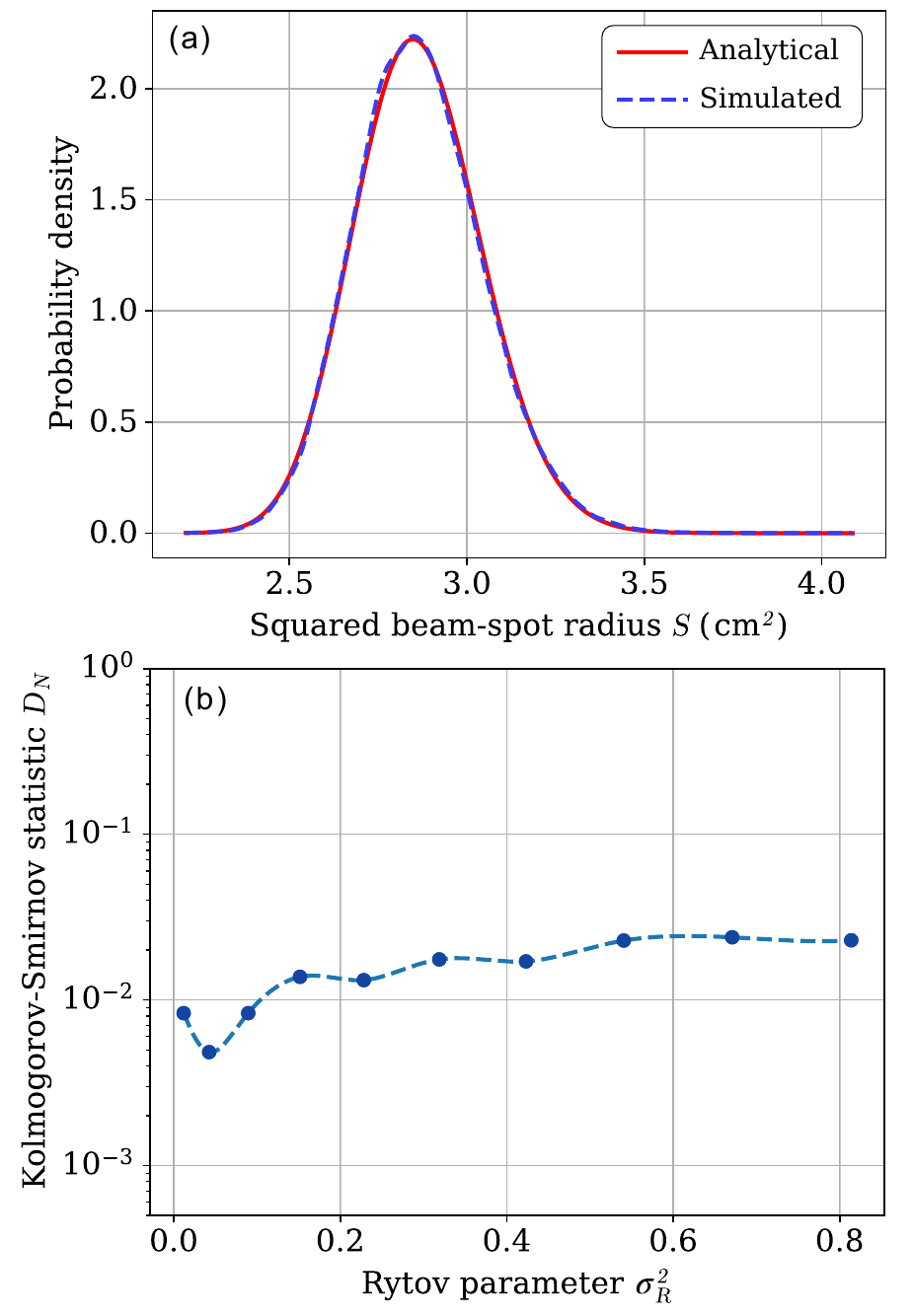}
		\caption{\label{Fig:distrS} Comparison between the log-normal and simulated probability distributions of $S$. (a) Solid and dashed lines show the probability distribution functions $P(S)$ for the log-normal distribution  (\ref{Eq:lognorm}) and the distribution estimated from the simulated data, respectively. (b) The Kolmogorov–Smirnov statistics (\ref{Eq:KS}) as a function of the Rytov parameter.
		 Here $L$ ranges from $500$ to $5000$~m, $\lambda = 808$~nm, $W_0=\sqrt{L\lambda / \pi}$, $C_n^2=10^{-15}$~m$^{-2/3}$,   $l_0=10^{-6}$~m, $L_0=5 \times 10^{3}$~m, $F_0=L$.}
	\end{figure}

Our first task is to validate the assumption that the probability distribution for $S$ can be well approximated by the log-normal distribution (\ref{Eq:lognorm}).  
To this end, we simulate the beam propagation through turbulent atmosphere by using the sparse-spectrum model  \cite{Charnotskii2013a,Charnotskii2013b,Charnotskii2020} for the phase-screen method \cite{Fleck1976,Frehlich2000,Lukin_book,Schmidt_book}.
Following the procedure detailed in Ref.~\cite{klen2023}, we obtain a sample $\{S_i|i=1\ldots N\}$ of $N$ realizations of the squared beam-spot radius $S$.
This sample is then used to estimate the parameters $\mu$ and $\sigma^2$ of the log-normal distribution.
The resulting parameters define the log-normal distribution, which we compare with the sampled distribution of $S$.
In particular, we directly reconstruct $P(S)$ by the smooth-kernel method and compare it with the related log-normal distribution.
Additionally, we estimate the empirical cumulative probability distribution function, 
	\begin{align}
		F_{N}(S)=\frac{1}{N}\sum_{i=1}^N \theta(S-S_i),
	\end{align}
where $\theta(S-S_i)$ is the Heaviside step function, and compare it to the cumulative distribution function of the log-normal distribution,
	\begin{align}
		F(S)=\int_{0}^{S}dS^\prime P(S^\prime)=\frac{1}{2}\erfc\left(\frac{\ln S-\mu}{\sigma\sqrt{2}}\right), 
	\end{align}
using the Kolmogorov–Smirnov statistics,
	\begin{align}\label{Eq:KS}
		D_N = \sup_{S\in[0+\infty)} \left| F_N(S) - F(S) \right|.
	\end{align}
Here $\erfc(x)$ denotes the complementary error function.
Our results show good agreement between the sampled data and the log-normal distribution for channels with weak impact of turbulence, i.e. with the Rytov parameter,
	\begin{align}\label{Eq:Rytov}
		\sigma_R^2=1.23C_n^2k^{7/6}L^{11/6},
	\end{align}
less than unity; see Fig.~\ref{Fig:distrS}.

Our next task consists in formulation of methods for evaluating three parameters of the circular-beam approximation based on the field correlation functions $\Gamma_2(\textbf{r};L)$ and $\Gamma_4(\textbf{r}_1, \textbf{r}_2;L)$.
While the beam-centroid coordinate $\sigma^2_{\text{bw}}$ can be determined from Eq.~(\ref{Eq:bwVariance}), the methods for determining the other two parameters, $\mu$ and $\sigma^2$, must be formulated separately.
A straightforward approach is to match them to the moments of the squared beam-spot radius, $\langle S\rangle$ and $\langle S^2\rangle$, as outlined in Appendix~\ref{App:MatchS}.
However, as shown in Ref.~\cite{klen2023} for the elliptic-beam approximation, this approach leads to a pronounced shift of the PDT mode (its maximum) across a wide range of aperture radii $a$.
Another problem is that beyond a limited domain of validity, the first two moments of the transmittance, $\left\langle \eta \right\rangle$ and $\left\langle \eta^2 \right\rangle$, are also biased.
The reason for this is caused by the fact that in realistic scenarios the beam profile is strongly non-Gaussian.
This implies that the beam-wandering, circular-beam, and elliptic-beam approximations are all affected by model misspecification bias.

In this section we address this problem by introducing an alternative method for evaluating the log-normal parameters $\mu$ and $\sigma^2$ in Eq.~(\ref{Eq:lognorm}).
Specifically, instead of matching them to the first two moments of the squared beam-spot radius, we match them to the first two moments of the transmittance, $\left\langle \eta \right\rangle$ and $\left\langle \eta^2 \right\rangle$, given by Eqs.~(\ref{Eq:etaAvrg}) and (\ref{Eq:eta2Avrg}), respectively.
The method is based on the fact that the first two moments of the circular-beam PDT (\ref{Eq:fullprob}) are given by
	\begin{align}
		&\left\langle \eta \right\rangle=\int_{0}^{+\infty}dS  P(S|\mu,\sigma^2)\left\langle\eta\right\rangle_S,\label{Eq:EqEta1}\\
		&\left\langle \eta^2 \right\rangle=\int_{0}^{+\infty}dS P(S|\mu,\sigma^2)\left\langle\eta^2\right\rangle_S.\label{Eq:EqEta2}
	\end{align}
Here, $P(S)\equiv P(S|\mu,\sigma^2)$ is the log-normal distribution (\ref{Eq:lognorm}) and $\langle\eta\rangle_S$ and $\langle\eta^2\rangle_S$ are the first two moments of transmittance conditioned on the beam-spot radius $S$, i.e., the moments of the PDT (\ref{Eq:BWPDT}).
Adapting results of Ref.~\cite{Esposito1967}, these moments can be expressed as
	\begin{align}
		\left<\eta\right>_S = 1 - \exp\!\left(-2\frac{a^2}{4\sigma^2_{\text{bw}} + S}\right),
	\end{align} 
	\begin{align}
		\left<\eta^2\right>_S& = 1-2 \exp\! \left(-2\frac{a^2}{4\sigma^2_{\text{bw}} + S}\right)\\& +  
		\exp\! \left(-\frac{\alpha^2}{2}\right)\left[1-Q\!\left(\frac{\alpha}{\sqrt{1-\beta^2}}, \frac{\alpha \beta}{\sqrt{1-\beta^2}}\right) \right.\nonumber
		\\
		&+ \left.Q\!\left(\frac{\alpha \beta}{\sqrt{1-\beta^2}}, \frac{\alpha}{\sqrt{1-\beta^2}}\right)\right],\nonumber
	\end{align}
where 
	\begin{align}
		\alpha=\frac{2a}{\sqrt{S}}\left[\frac{2 p(p+1)}{2 p^2+3 p+1}\right]^{1 / 2}, 
	\end{align} 
	\begin{align}
		\beta = (2p+1)^{-1}, \quad p = \frac{1}{8}\frac{S}{\sigma^2_{\text{bw}}},
	\end{align} 
and $Q(x,y)$ is the Marcum $Q$ function of the first order; see Refs.~\cite{Marcum_book,Agrest_book,Vasylyev2013}.

The transmittance–moment matching method for the circular-beam approximation can now be summarized as follows.
First, we calculate the parameters $\sigma_{\mathrm{bw}}^2$, $\langle \eta \rangle$, and $\langle \eta^2 \rangle$ using Eqs.~(\ref{Eq:bwVariance}), (\ref{Eq:etaAvrg}), and (\ref{Eq:eta2Avrg}), respectively.
Second, we treat Eqs.~(\ref{Eq:EqEta1}) and (\ref{Eq:EqEta2}) as a system of algebraic equations for $\mu$ and $\sigma^2$, and solve it numerically via least-squares optimization.
As an initial guess one can use the values of $\mu$ and $\sigma^2$ used in the method of matching the moments of $S$, described in Appendix \ref{App:MatchS}; see Eqs.~(\ref{Eq:mu}) and (\ref{Eq:sigma}).\footnote{In practice, we also constrain the values of $\mu$ and $\sigma^2$. Specifically, we require the mean of the corresponding $S$ distribution to lie within a factor of 5 of its initial guess. Additionally, we require that the probability of an instantaneous value of $S$ exceeding 10 times its mean is less than 1\%. For the considered case, this leads to the constraint $10^{-6} < \sigma^2 < 2$.}
The resulting values of $\mu$ and $\sigma^2$ are then used in Eqs.~(\ref{Eq:fullprob}) and (\ref{Eq:lognorm}) to evaluate the PDT.
Since $\langle \eta \rangle$ and $\langle \eta^2 \rangle$ are now directly matched, their values are no longer biased.

In realistic applications, constant (nonfluctuating) losses---caused by absorption, scattering, and imperfections in the transmitter and receiver optical systems---are always present, in addition to the fluctuating losses induced by atmospheric turbulence. 
These constant losses may be naturally incorporated into the transmittance–moment matching method. 
To do so, one replaces the left-hand sides of Eqs.~(\ref{Eq:EqEta1}) and (\ref{Eq:EqEta2}) with $\eta_{\mathrm{c}}\langle\eta\rangle$ and  $\eta_{\mathrm{c}}^2\langle\eta^2\rangle$, respectively, where $\eta_{\mathrm{c}}$ represents the efficiency associated with the constant losses.
In other words, the modified left-hand sides of these equations correspond to realistic transmittance moments that account for all experimental losses. 
Within this framework, the constant losses are effectively modeled by adjusting the parameters $\mu$ and $\sigma^2$ of the log-normal distribution for $S$. 
This approach represents an optional modification of the described method, applicable in suitable scenarios.


\section{Phase approximation of the Huygens–Kirchhoff method}
\label{Sec:Analytical}

In the circular-beam approximation with transmittance matching, the parameters $\left\langle\eta\right\rangle$, $\left\langle\eta^2\right\rangle$, and $\sigma_{\mathrm{bw}}^2$ are required; they are determined by the second- and fourth-order field correlation functions.
Several methods have been proposed in the literature to analytically calculate these functions; see, e.g., Refs.~\cite{Tatarskii,Tatarskii2016,Fante1975,Fante1980,Andrews_book,Berman2007}.
In this paper we adopt the phase approximation of the Huygens–Kirchhoff method \cite{Aksenov1979,Banakh1979,Mironov1977}, which has already been applied in the context of the elliptic-beam approximation \cite{vasylyev16}.

The field correlation functions in this approximation for a focused beam ($L=F$) are given by (see, e.g., Supplemental Materials in Ref.~\cite{vasylyev16})
    \begin{align}\label{Eq:Gamma2approx}
		\Gamma_2 \left(\textbf{r}; L\right) = \frac{k^2}{4 \pi^2 L^2}  \int_{\mathds{R}^2} d^2\textbf{r}^\prime &\exp \bigg[-\frac{|\textbf{r}^\prime|^2}{2 W_0^2}\\
		&-2i\textbf{r} \cdot \textbf{r}^\prime\frac{\Omega}{W_0^2}-\frac{1}{2}D_s\left(0, \textbf{r}^\prime\right) \bigg],\nonumber
	\end{align}
	\begin{align}\label{Eq:Gamma4approx}
		&\Gamma_4 \left(\textbf{r}_1, \textbf{r}_2; L\right) = \frac{k^4}{4 \pi^5 L^4 W_0^2}  \int_{\mathds{R}^6} d^2\textbf{r}_1^\prime d^2\textbf{r}_2^\prime d^2\textbf{r}_3^\prime \\
		&\exp \bigg[ -\frac{1}{W_0^2}\left(|\textbf{r}_1^\prime|^2 + |\textbf{r}_2^\prime|^2 + |\textbf{r}_3^\prime|^2\right)+2i\textbf{r}_1^\prime \cdot \textbf{r}_2^\prime\frac{\Omega}{W_0^2}\nonumber \\
		& -2i(\textbf{r}_1-\textbf{r}_2)\cdot\textbf{r}_2^\prime\frac{\Omega}{W_0^2} -2i(\textbf{r}_1+\textbf{r}_2)\cdot\textbf{r}_3^\prime\frac{\Omega}{W_0^2}  \bigg]\nonumber\\
		&\times \exp \bigg[ \frac{1}{2} \sum_{j=1,2} \big\{D_s(\textbf{r}_1-\textbf{r}_2, \textbf{r}_1^\prime+(-1)^j\textbf{r}_2^\prime)\nonumber \\
		&-D_s(\textbf{r}_1-\textbf{r}_2, \textbf{r}_1^\prime+(-1)^j\textbf{r}_3^\prime) -D_s(0, 	\textbf{r}_2^\prime+(-1)^j\textbf{r}_3^\prime)\big\} \bigg].\nonumber
	\end{align}
Here
    \begin{align}\label{Eq:structureF}
		D_s(\textbf{r},\textbf{r}^\prime)=2\rho_0^{-5/3} \int_0^1 d\xi |\textbf{r}\xi+\textbf{r}^\prime(1-\xi)|^{5/3}
	\end{align}	
is the phase structure function,
    \begin{align}\label{Eq:rho}
		\rho_0=(1.46C_n^2k^2L)^{-5/3}
	\end{align}
is the spatial coherence radius, and $\Omega={kW_0^2}/{2L}$ is the Fresnel number.

\begin{figure*}[ht!]
	\centering
	\includegraphics[width=0.8\linewidth]{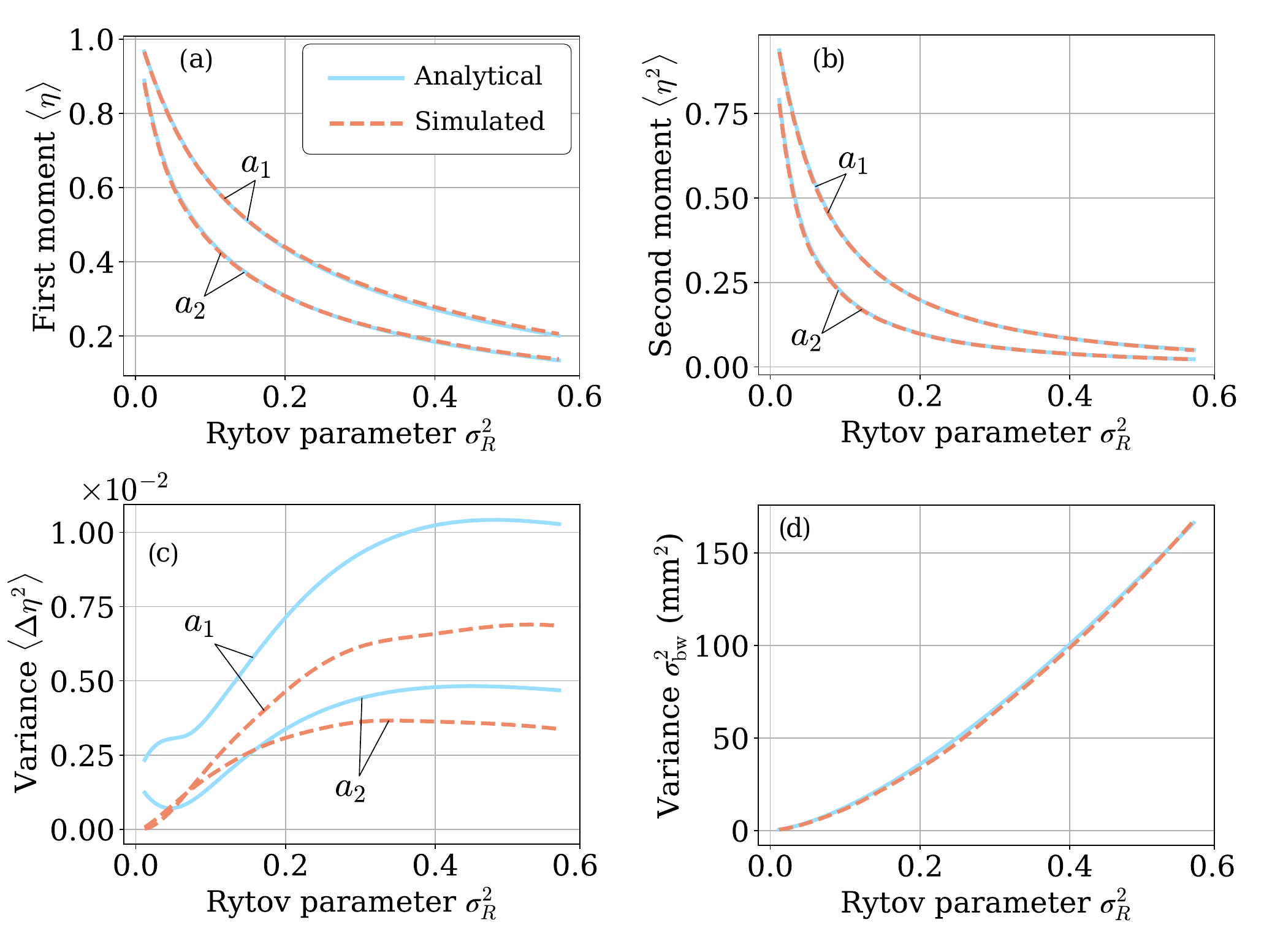}
	\caption{\label{Fig:sim-analy} Comparison of analytical results obtained using the Huygens-Kirchhoff method (solid blue lines) with simulated data based on the sparse-spectrum model of the phase-screen method (dashed orange lines) for (a)~$\langle \eta \rangle$, (b)~$\langle \eta^2 \rangle$, (c)~$\langle \Delta\eta^2 \rangle$, and (d)~$\sigma_{\mathrm{bw}}^2$ .
		The channel parameters are the same as in Fig.~\ref{Fig:distrS}.
		The propagation distance $L$ ranges from $500$~m to $4000$~m.
		In plots~(a)–(c), the aperture radii are $a_1=15$~mm, $a_2=12$~mm.
	}        
\end{figure*}

To obtain the variance $\sigma_{\mathrm{bw}}^2$, we substitute the above expressions into Eq.~(\ref{Eq:bwVariance}). 
For the channels with weak impact of the turbulence, we expand the obtained relations with respect to $\rho^{-5/3}$ up to the second order.
Using approximations suitable for the weak–turbulence regime and carrying out the integration (see Appendix~\ref{App:AnalyticalFormulas}), we arrive at
    \begin{align}\label{Eq:bwFinal}
	\sigma_{\mathrm{bw}}^2=0.31 W_0^2 \sigma_R^2\Omega^{-7/6}-0.06  W_0^2 \sigma_R^4\Omega^{-1/3}.
	\end{align}
Here $\sigma_R^2$ is the Rytov parameter, cf. Eq.~(\ref{Eq:Rytov}).

A similar technique yields approximate analytical expressions for $\langle\eta\rangle$ and $\langle\eta^2\rangle$.
In this case the analytical forms of the field correlation functions, Eqs.~(\ref{Eq:Gamma2approx}) and (\ref{Eq:Gamma4approx}),  are substituted into Eqs.~(\ref{Eq:etaAvrg}) and (\ref{Eq:eta2Avrg}). 
Applying approximations valid in the regime of weak–turbulence, see Appendix~\ref{App:AnalyticalFormulas}, yields analytical expressions for these moments: 
    \begin{align}\label{Eq:etaFinal}
		\left\langle\eta\right\rangle =1-\exp\left( -\frac{a^2}{0.5 W_0^2\Omega^{-2}+0.66 W_0^2\sigma_R^2\Omega^{7/6}} \right),
	\end{align}
	\begin{multline}\label{Eq:eta2Final}
		\left\langle\eta^2\right\rangle =\left[ 1-\exp\left( -\frac{4a^2}{W_0^2\Omega^{-2}\left( 	1+2v\Omega^2 \right)} \right)\right]\\
	\times\left[ 1-\exp\left( -\frac{a^2 \left( 1+2v\Omega^2 \right)}{vW_0^2} \right)\right].
	\end{multline}
Here $v=\Omega^{-2}+3.17\sigma_R^2\Omega^{-7/6}$.
It is also important to note that these expressions are confirmed to be valid only for the case $W_0=\sqrt{L\lambda / \pi}$.

To verify the validity of these analytical expressions, we compare them with simulation results obtained using the sparse-spectrum model of the phase-screen method, following the approach of Ref.~\cite{klen2023}.
The comparison is presented in Fig.~\ref{Fig:sim-analy} for channels of different length $L$ plotted as a function of the Rytov parameter $\sigma_{\mathrm{R}}^2$ [see Eq.~(\ref{Eq:Rytov})]. 
We observe good agreement between the results obtained via the Huygens–Kirchhoff and phase-screen method for the variance of the beam-centroid coordinate $\sigma_{\mathrm{bw}}^2$.
With regard to the transmittance–moment matching method, we find good agreement for the first and second moments of transmittance, $\langle \eta\rangle$ and  $\langle \eta^2\rangle$.
However, in this case the transmittance variance $\langle \Delta\eta^2\rangle$ plays a crucial role and its values are much smaller compared to the values of moments. 
As shown, the agreement for $\langle \Delta\eta^2\rangle$ is significantly worse in several cases. 

The proposed method of matching to the transmittance moments also requires an initial guess in the iterative optimization procedure used to determine the parameters $\mu$ and $\sigma^2$.
These initial values can be obtained by matching them to the moments of the squared beam-spot radius, $\langle S\rangle$ and $\langle S^2\rangle$, as given by Eqs.~(\ref{Eq:mu}) and (\ref{Eq:sigma}).
Analytical expressions for the latter can be derived using the phase approximation of the Huygens–Kirchhoff method, as discussed in Appendix~\ref{App:AnalyticalFormulas}.

The phase approximation of the Huygens–Kirchhoff method can also be applied in the regime of strong turbulence \cite{Aksenov1979,Banakh1979,Mironov1977,vasylyev16}. 
However, in this case the derived parameters suffer from even greater inaccuracies. 
Moreover, as shown in Ref.~\cite{klen2023}, the validity of analytical models depends only weakly on the turbulence strength, while the aperture radius plays a much more significant role. 
For these reasons we restrict our analysis to the case of weak turbulence, focusing on assessing the validity of the model itself.


\section{Examples and model validation}
\label{Sec:Examples}

In this section we demonstrate the performance of our model across different scenarios. 
In addition to the transmittance-matching technique, we also consider matching to the moments of $S$ to clearly identify the validity domain of both methods.
The parameters $\sigma_{\mathrm{bw}}^2$, $\langle \eta \rangle$, and $\langle \eta^2 \rangle$ in the former case, and $\sigma_{\mathrm{bw}}^2$, $\langle S\rangle$, and $\langle S^2 \rangle$ in the latter case, are evaluated using the phase approximation of the Huygens–Kirchhoff method.
However, as discussed in the previous section, this approximation itself introduces errors.
As a result, it can be difficult to determine whether discrepancies between analytical and numerically simulated PDTs arise from the circular-beam model itself or from inaccuracies in the parameter evaluation.
In the latter case, such discrepancies could, in principle, be reduced by employing more accurate methods of parameter evaluation.

This issue can be resolved by using numerically simulated values of the parameters $\sigma_{\mathrm{bw}}^2$, $\langle \eta \rangle$, $\langle \eta^2 \rangle$, $\langle S\rangle$, and $\langle S^2 \rangle$.
Although this approach cannot be considered a practical tool for PDT evaluation, these values serve as a kind of ground truth for benchmarking the best achievable analytical parameter estimates.
If discrepancies persist in this case, it clearly indicates that the model itself fails in the corresponding domain.
Conversely, if the numerically determined parameters yield good agreement, this suggests that the circular-beam model remains valid and that further refinement of the analytical evaluation of $\sigma_{\mathrm{bw}}^2$, $\langle S \rangle$, $\langle S^2 \rangle$, $\langle \eta \rangle$, and $\langle \eta^2 \rangle$ could lead to even better results.

\begin{figure}[ht!]
       \centering
        \includegraphics[width=1\linewidth]{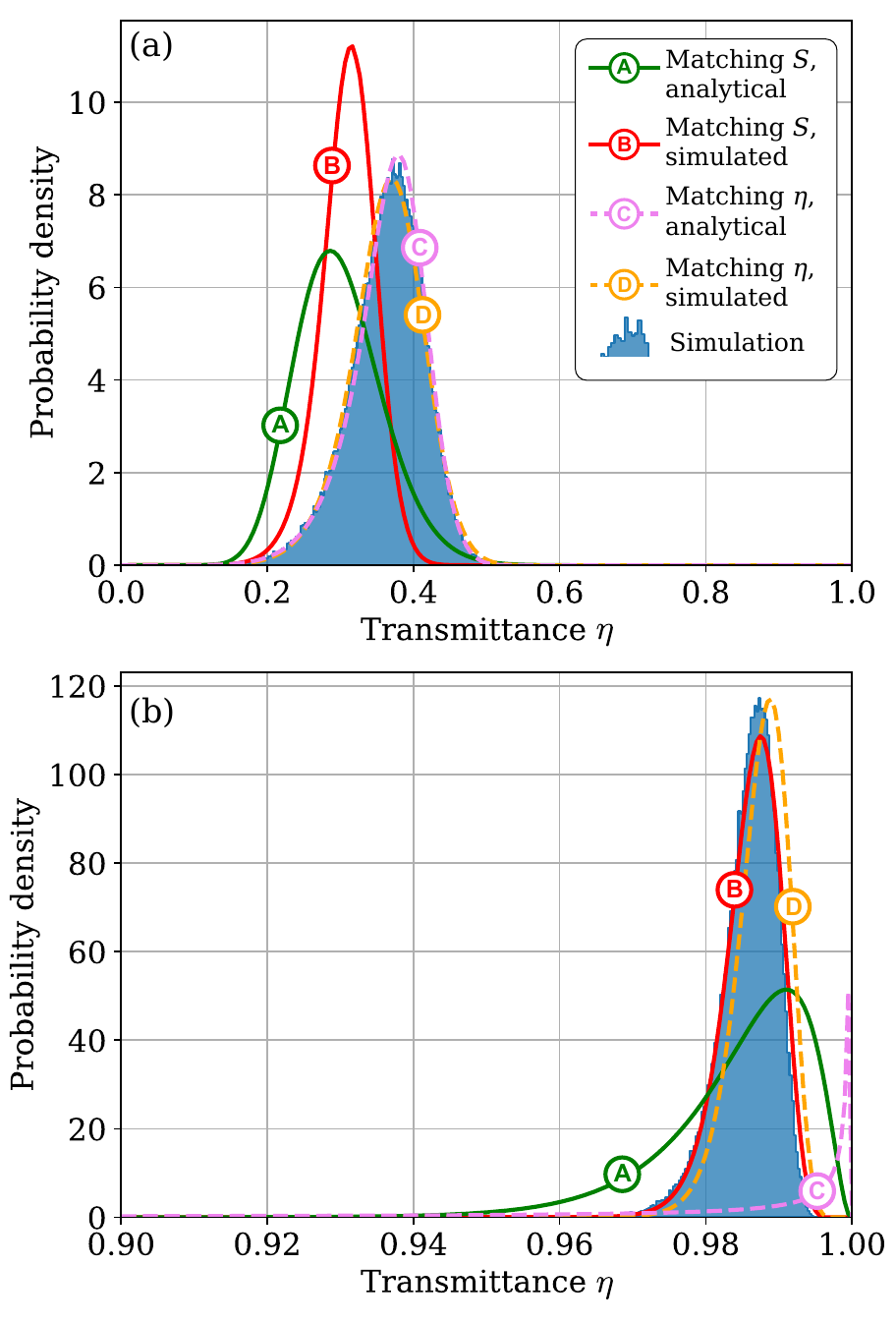}
        \caption{\label{Fig:PDT_distrib} Comparison of simulated PDT histograms with PDTs obtained using the circular-beam approximation, based on different matching methods and either analytical or numerically simulated parameters.
       	The channel parameters are the same as in Fig.~\ref{Fig:distrS}, with (a) $L = 2000$~m, $a = 12$~mm and (b) $L = 1000$~m, $a = 25$~mm.
       	The green solid line (A) and the red solid line (B) represent the method of matching the moments of $S$ using analytical and numerically simulated values, respectively.
       	The violet dashed line (C) and the orange dashed line (D) correspond to the transmittance–moment matching method, also using analytical and simulated values, respectively.
        } 
    \end{figure}

Let us start with an example characterized by an aperture radius $a$ for which the mode of the PDT, obtained using the circular-beam approximation with matching the moments of $S$, significantly deviates from that of the numerically simulated PDT. 
A similar discrepancy is observed when using the beam-wandering and elliptic-beam approximations, rendering their application in this regime highly inaccurate; see Ref.~\cite{klen2023}.
The PDTs for such a channel are shown in Fig.~\ref{Fig:PDT_distrib}(a).
Lines A and B, obtained using this method, clearly deviate from the numerically simulated histograms.
In contrast, the transmittance–moment matching method yields a markedly improved result: the corresponding lines C and D show good agreement with the histogram, both for analytically calculated and numerically simulated parameter values.

Another example concerns the aperture radius $a$ for which the Kolmogorov–Smirnov statistic indicates the best agreement between analytical and numerical data for the beam-wandering and elliptic-beam approximations, as reported in Ref.~\cite{klen2023}.
The similar situation in this case is observed for the circular-beam approximation.
An additional feature of this example is the close match between the analytically calculated and numerically simulated moments of $S$.
The corresponding PDTs are shown in Fig.~\ref{Fig:PDT_distrib}(b).
For simulated parameter values, both matching methods (lines B and D) yield good agreement.
However, when analytical parameter values are used, the method based on matching the moments of $S$ (line A) performs significantly better than the transmittance–moment matching method (line C).   

\begin{figure}[ht!]
	\centering
	\includegraphics[width=1\linewidth]{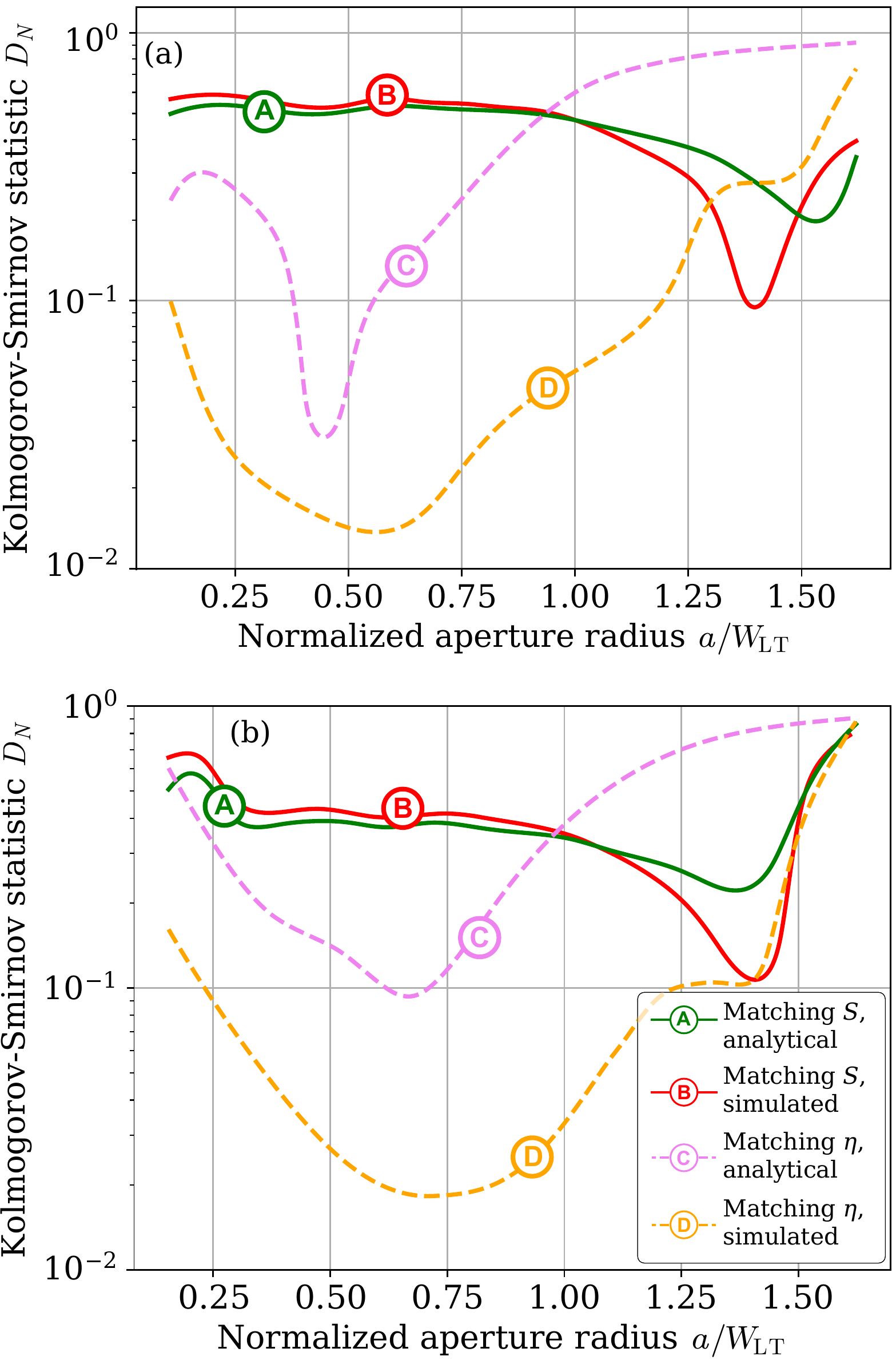}
	\caption{\label{Fig:PDT_KS} Kolmogorov–Smirnov statistics quantifying the difference between numerically simulated PDTs and analytical PDTs for (a) $L = 2000$~m with $W_{\text{LT}} = 28$~mm and aperture radius $a$ ranging from 3 to 47~mm, and (b) $L = 1000$~m with $W_{\text{LT}} = 19$~mm and $a$ ranging from 3 to 32~mm.
	The lines correspond to the cases shown in Fig.\ref{Fig:PDT_distrib}.
	For the remaining channel parameters, see Fig.\ref{Fig:distrS}.
	} 
\end{figure}

To summarize these results, we consider the Kolmogorov–Smirnov statistic comparing the PDT obtained using the circular-beam approximation with numerically simulated data (see Fig.~\ref{Fig:PDT_KS}).
The transmittance–moment matching method shows good agreement with the simulated data across a wide range of aperture radii $a$.
However, in the range where the aperture radius $a$ lies between $1.25W_{\mathrm{LT}}$ and $1.50W_{\mathrm{LT}}$, where
	\begin{align}
		W_{\mathrm{LT}}^2=4\int_{\mathds{R}^2} d^2\textbf{r} x^2 \Gamma_2(\textbf{r};L)
	\end{align} 
is the long-term beam radius, the method of matching the moments of $S$ is preferable---particularly when using analytical expressions derived from the Huygens–Kirchhoff method.

\begin{figure}[h!]
	\centering
	\includegraphics[width=1\linewidth]{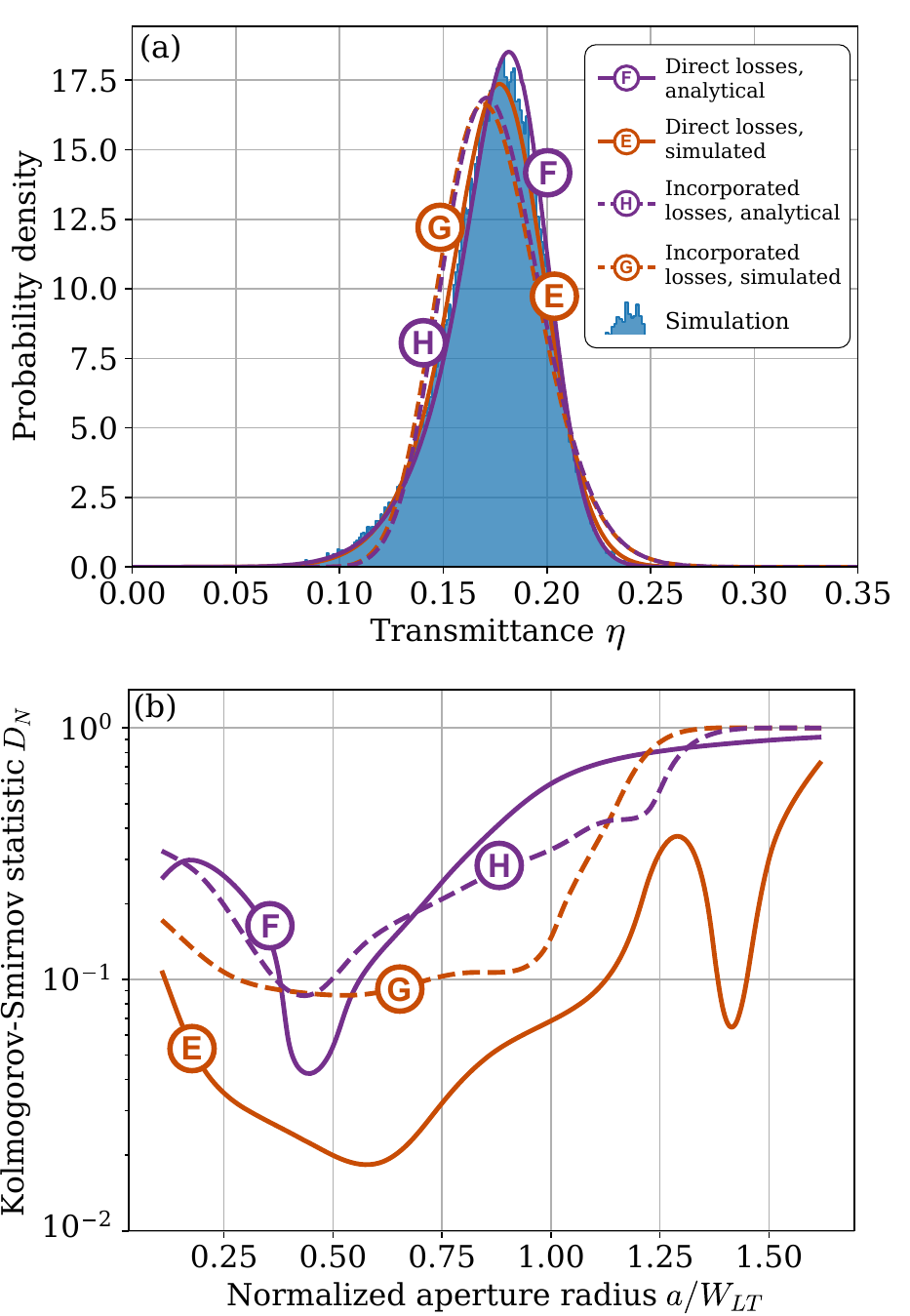}
	\caption{\label{Fig:PDT_distrib+KS_determ_losses} 
		Comparison of two approaches for incorporating constant losses into the transmittance-moment matching method.
		Solid lines F (violet) and E (brown) correspond to the direct rescaling of the PDT using analytical and numerically simulated moments of $\eta$, respectively.
		Dashed lines H (violet) and G (brown) represent the indirect incorporation of constant losses into the parameters of the log-normal distribution for $S$.
		Panel (a) shows analytical PDTs and corresponding numerically simulated histograms for $L=2000$~m, $a=12$~mm.
		Panel (b) presents the corresponding Kolmogorov–Smirnov statistics for $L=2000$~m, $W_{\text{LT}}=28$~mm, and aperture radius $a$ ranging from $3$~mm to $47$~mm.    
		For the rest of the channel parameters, see Fig.~\ref{Fig:distrS}.
		The values of constant losses are specified in the text.} 
\end{figure}

Let us examine whether incorporating constant losses into the parameters of the log-normal distribution for $S$---as discussed in the final paragraph of Sec.~\ref{Sec:Circbeam}---yields appropriate results.
To this end, we consider PDTs obtained using two variants of the transmittance–moment matching method. 
The first variant assumes that constant losses result in a rescaling of the PDT as $\mathcal{P}_{\mathrm{loss}}(\eta)=\eta_{\mathrm{c}}^{-1}\mathcal{P}(\eta/\eta_{\mathrm{c}})$.
The second variant involves replacing the left-hand sides of Eqs.~(\ref{Eq:EqEta1}) and (\ref{Eq:EqEta2}) with $\eta_{\mathrm{c}}\langle\eta\rangle$ and  $\eta_{\mathrm{c}}^2\langle\eta^2\rangle$, respectively.
The results are presented in Fig.~\ref{Fig:PDT_distrib+KS_determ_losses}.
We assume constant losses of $3$~dB from the optical systems of the transmitter and receiver, and $0.1$~dB/km due to atmospheric absorption and scattering.
As can be seen, incorporating constant losses into the parameters of the log-normal distribution for $S$ generally does not lead to any improvement.

\begin{figure}[ht!]
	\centering
	\includegraphics[width=1\linewidth]{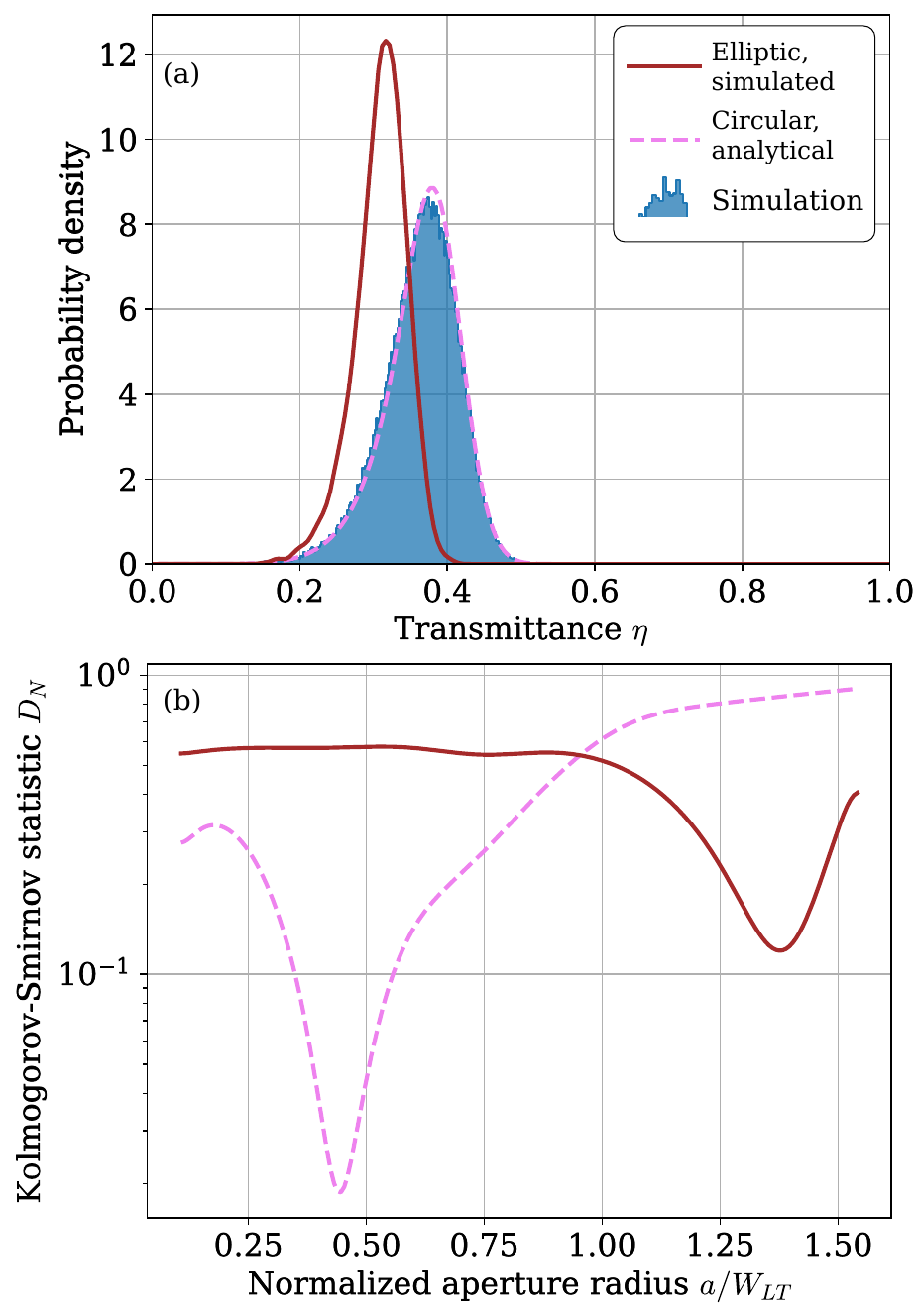}
	\caption{\label{Fig:Elliptic} The elliptic-beam approximation [solid line (A)] obtained with the numerically simulated parameters and the circular-beam approximation with the analytical parameters $\langle \eta \rangle$ and $\langle \eta^2 \rangle$ [dashed line (B)] are shown.
			Panel (a) demonstrates the PDT for the same channel and aperture radius as in Fig.~\ref{Fig:PDT_distrib}(a).
			Panel (b) presents the corresponding Kolmogorov–Smirnov statistics as a function of the aperture radius $a$.
	} 
\end{figure}

As discussed above, the circular-beam approximation makes it possible to eliminate the model misspecification bias by matching the parameters $\mu$ and $\sigma^2$ to $\langle \eta \rangle$ and $\langle \eta^2 \rangle$ instead of $\langle S\rangle$ and $\langle S^2 \rangle$.
Applying a similar procedure to the elliptic-beam approximation, although in principle feasible, is considerably more challenging.
First, this model involves an additional parameter in the two-dimensional log-normal distribution, which accounts for the correlation between the ellipse semiaxes. As a result, three parameters must be determined from only two equations, making the system not fully determined.
Second, the elliptic-beam approximation requires more elaborate numerical techniques, which substantially complicates the optimization procedure.

As shown in Fig.~\ref{Fig:Elliptic}, the circular-beam approximation combined with the transmittance-matching method exhibits significantly better agreement with numerical simulations than the elliptic-beam approximation over a broad range of aperture radii $a$.
Importantly, this comparison is performed using numerically simulated (i.e., best achievable) parameters for the elliptic-beam approximation and analytical parameter values for the circular-beam approximation.
These results clearly demonstrate that, for suitable aperture sizes, matching transmittance moments within the circular-beam framework provides a more efficient and accurate parameter estimation technique than the standard elliptic-beam approach.


%
\section{Application: Free-space transfer of nonclassical optical properties}
\label{Sec:Application}

In this section we demonstrate how the developed PDT model can be applied to analyze nonclassical properties of quantum light transmitted through a free-space channel.
In quantum optics, the quantum state of a light mode is considered nonclassical \cite{titulaer65,mandel86,mandel_book,vogel_book,agarwal_book,Schnabel2017,sperling2018a,sperling2018b,sperling2020} if it cannot be represented as a statistical mixture of coherent states.
In other words, its Glauber-Sudarshan $P$ function, $P(\alpha)$, fails to have the interpretation of a classical probability distribution.
Nonclassicality can be revealed from measurement statistics---for example, through photocounting or balanced homodyne detection.

Let us start with photocounting measurements performed using a photon-number resolving detector.
In this case a sufficient condition of nonclassicality is the negativity of the Mandel $Q$ parameter,
	\begin{align}\label{Eq:Mandel}
	 	Q	=\frac{\left\langle \Delta n^2 \right\rangle}{\left\langle n \right\rangle}-1,
	\end{align}
which indicates a sub-Poissonian character of the photon-number statistics \cite{mandel79,mandel_book}.
Here, $\left\langle n \right\rangle$ and $\left\langle \Delta n^2 \right\rangle$ denote the mean and the variance of the photon number, respectively. 
The Mandel $Q$ parameters of the input and output modes---$Q_{\mathrm{in}}$ and $Q_{\mathrm{out}}$, respectively---are related for a free-space channel as
	 \begin{align}\label{Eq:MandelInOut}
	  	Q_{\text{out}}=\frac{\left\langle \eta^2\right\rangle}{\left\langle \eta\right\rangle}Q_{\text{in}}+\frac{\left\langle \Delta \eta^2 \right\rangle}{\left\langle \eta \right\rangle} \left\langle n \right\rangle_{\text{in}},
	 \end{align} 
where $\left\langle n \right\rangle_{\text{in}}$ is the mean photon number at the input; see Refs.~\cite{semenov09,semenov2018}.
Throughout this section we incorporate all constant losses into the model by treating the transmittance $\eta$ as the product $\eta_{\mathrm{c}}\eta$.    
 
Equation~(\ref{Eq:MandelInOut}) depends only on two first moments of $\eta$.
Therefore, the transmittance-matching moments method is ideally suited for describing the transfer of sub-Poissonian photon statistics.
Any discrepancies in this case can arise only from imperfections in the analytical approximations used to compute these moments. 

\begin{figure}[ht!]
	\centering
	\includegraphics[width=1\linewidth]{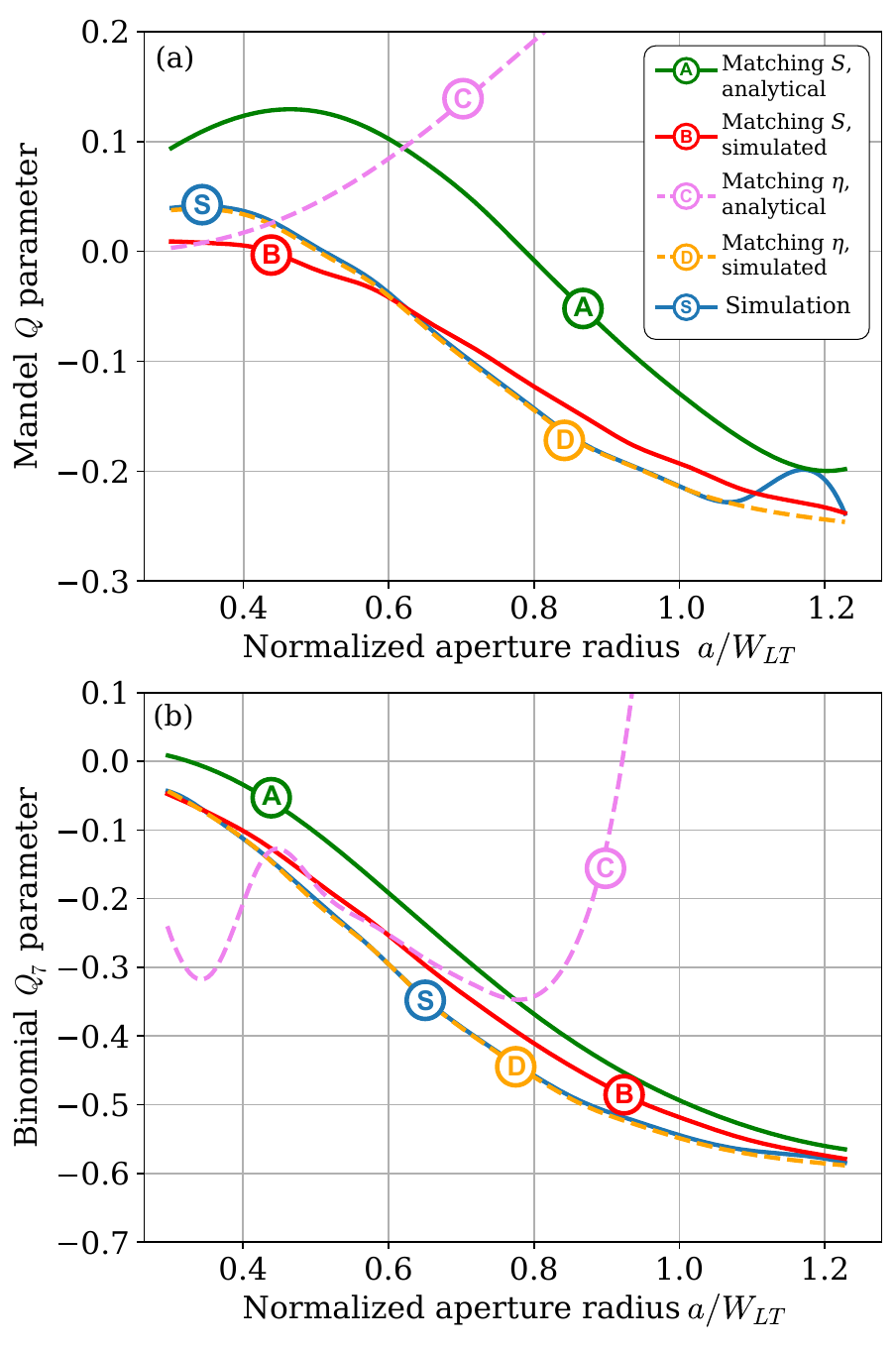}
	\caption{\label{Fig:app_Mandel+Binomial} 
		Comparison of Mandel $Q$ parameter and binomial $Q_7$ parameter calculated by means of simulated PDT and PDTs obtained using the circular-beam approximation, based on different matching methods and either analytical or numerically simulated parameters.
		The blue solid line (S) stands for using simulated PDT: for the correspondence of other lines, see Fig.~\ref{Fig:PDT_distrib}.
		The channel parameters are the same as in Fig.~\ref{Fig:PDT_distrib}(a), with aperture radius $a$ ranging from $8$ to $34$~mm, with $\chi=0.4$, $\alpha_0=6$, and $\eta_\text{c}=0.48$.
	} 
\end{figure}

To be more specific, we consider propagation of amplitude-squeezed coherent states $\ket{\alpha_0, \chi }=\hat{D}(\alpha_0)\hat{S}(\chi)\ket{0}$, where $\hat{D}(\alpha_0)$ and $\hat{S}(\chi)$ are the displacement operator and squeezing operator, respectively, $\chi>0$ is the squeezing parameter, $\alpha_0$ is the coherent amplitude, and $\ket{0}$ is the vacuum state.
In Fig.~\ref{Fig:app_Mandel+Binomial}(a) we present the Mandel $Q$ parameter as a function of aperture radius for different parameter-matching techniques, along with numerically simulated data (line S).
The transmittance–moment matching method using simulated moments (line D) reproduces the numerical results exactly. 
However, the same method using analytical moment values (line C) shows a significant discrepancy.
This is due to the second term in Eq.~(\ref{Eq:MandelInOut}): since $\left\langle n \right\rangle_{\text{in}}$ is large for high values of $\alpha_0$, even a small error in the analytical expressions for $\left\langle \Delta \eta^2 \right\rangle$ can lead to a large deviation in $Q_{\text{out}}$.
Remarkably, matching the moments of $S$ provides a reliable result: it yields only a small discrepancy for simulated moments of $S$ (line B) and a qualitatively correct result when using analytical values (line A).   
 
In a more realistic scenario, one can consider so-called click detectors.
The corresponding measurement scheme is based on balanced spatial \cite{paul1996,castelletto2007,schettini2007,blanchet08} or temporal \cite{achilles03,fitch03,rehacek03} splitting of a light beam into $N$ parts, each detected by an on–off detector.
The number of triggered detectors, $n$, corresponds to the number of clicks, which approaches the actual photon number in the limit $N\rightarrow+\infty$.
The click-number distribution at the output is given by
	\begin{align}
		\mathcal{P}_n=\int_{\mathbb{C}}d^2\alpha P_{\mathrm{out}}(\alpha)\Pi_n(\alpha),
	\end{align}
where $P_{\mathrm{out}}(\alpha)$ is the Glauber–Sudarshan $P$ function of the output mode, related to $P_{\mathrm{in}}(\alpha)$ via Eq.~(\ref{Eq:InOut}).
The function $\Pi_n(\alpha)$ represents the positive operator-valued measure element in $Q$ representation corresponding to $n$ clicks and is given by \cite{sperling12a}
	\begin{align}\label{Eq:POVMCLICK}
		\Pi_n(\alpha)=\binom{N}{n} \exp&\left[ -\frac{|\alpha|^2(N-n)}{N}\right]\\
	&\times \left[1-\exp\left(-\frac{|\alpha|^2}{N}\right)\right]^n \nonumber ,
	\end{align}
see Ref.~\cite{sperling12a}.
To witness nonclassicality using this detection scheme, one can use the binomial $Q$ parameter \cite{sperling12c}:
	\begin{align}\label{Eq:Binomial}
		Q_N=N\frac{\left\langle \Delta n^2\right\rangle}{\left\langle n \right\rangle\left(N-\left\langle n \right\rangle\right)} -1 \, .
	\end{align}	
Importantly, unlike the Mandel $Q$ parameter, it is not possible to express an input–output relation for $Q_N$ similar to Eq.~(\ref{Eq:MandelInOut}). 
The output binomial parameter depends on the full PDT, not just its first two moments.

Let us consider the case $N=7$ and the same state as in the previous example---in this regime, the click detector operates well below saturation.
Figure~\ref{Fig:app_Mandel+Binomial}(b) shows the dependence of the binomial parameter $Q_7$ on the aperture radius. 
The transmittance–moment matching method using the simulated  moments (line D)  again demonstrates good agreement with the simulated data (line S), indicating that the circular-beam PDT accurately captures information about the numerical PDT beyond just the first two moments of $\eta$.
The same method with analytical moments also performs well within the domain $0.43\lesssim a/W_{\mathrm{LT}}\lesssim 0.8$.
Reasonable results are obtained with matching the moments of $S$ as well, where the outcome with simulated moments (line B) surpasses that with analytical moments (line A).
     
Another nonclassical effect---quadrature squeezing---requires measurement of a field quadrature, e.g., $\hat{x}=2^{1/2}\left(\hat{a}+\hat{a}^\dagger \right)$, where $\hat{a}$ is the field annihilation operator.
The corresponding measurement procedure---balanced homodyne detection---can be adapted for free-space channels by sending the local oscillator in the same spatial mode as the signal but with orthogonal polarization \cite{elser09,heim10,semenov12}.
This configuration effectively suppressed the dephasing effect, making it negligibly small.
 
The quadrature variance can be expressed in terms of the normally-ordered quadrature variance $ \left\langle : \Delta \hat{x}^2 : \right\rangle$ as  
	\begin{align}\label{Eq:Quadrat}
		\left\langle \Delta \hat{x}^2 \right\rangle = \left\langle : \Delta \hat{x}^2 : \right\rangle +\frac{1}{2}.
	\end{align}
Since for classical light the normally ordered variance is always non-negative, the condition $\left\langle \Delta \hat{x}^2 \right\rangle<1/2$ serves as a witness of quadrature squeezing, which is an example of nonclassical behavior.
The corresponding input–output relation is given by \cite{semenov12,semenov2018}
	\begin{align}\label{Eq:QuadratInOut}
\left\langle : \Delta \hat{x}^2 : \right\rangle_\text{out} = \left\langle \eta  \right\rangle \left\langle : \Delta \hat{x}^2 : \right\rangle_\text{in} + \left\langle  \Delta T^2  \right\rangle \left\langle \hat{x} \right\rangle_\text{in}^2 \, ,
\end{align}
where $T=\sqrt{\eta}$ is the transmission coefficient.
Therefore, the output squeezing depends not only on the first moment $\langle\eta\rangle$, but also on the variance $\left\langle  \Delta T^2  \right\rangle$, which involves $\left\langle \sqrt{\eta}  \right\rangle$.
This fractional moment is not accounted for in the transmittance–moment matching method by Eqs.~(\ref{Eq:EqEta1}) and (\ref{Eq:EqEta2}).
Hence, it is important to verify whether the circular-beam PDT with the transmittance-moment matching method can adequately describe the transmission of quadrature squeezing.

\begin{figure}[ht!]
       \centering
        \includegraphics[width=1\linewidth]{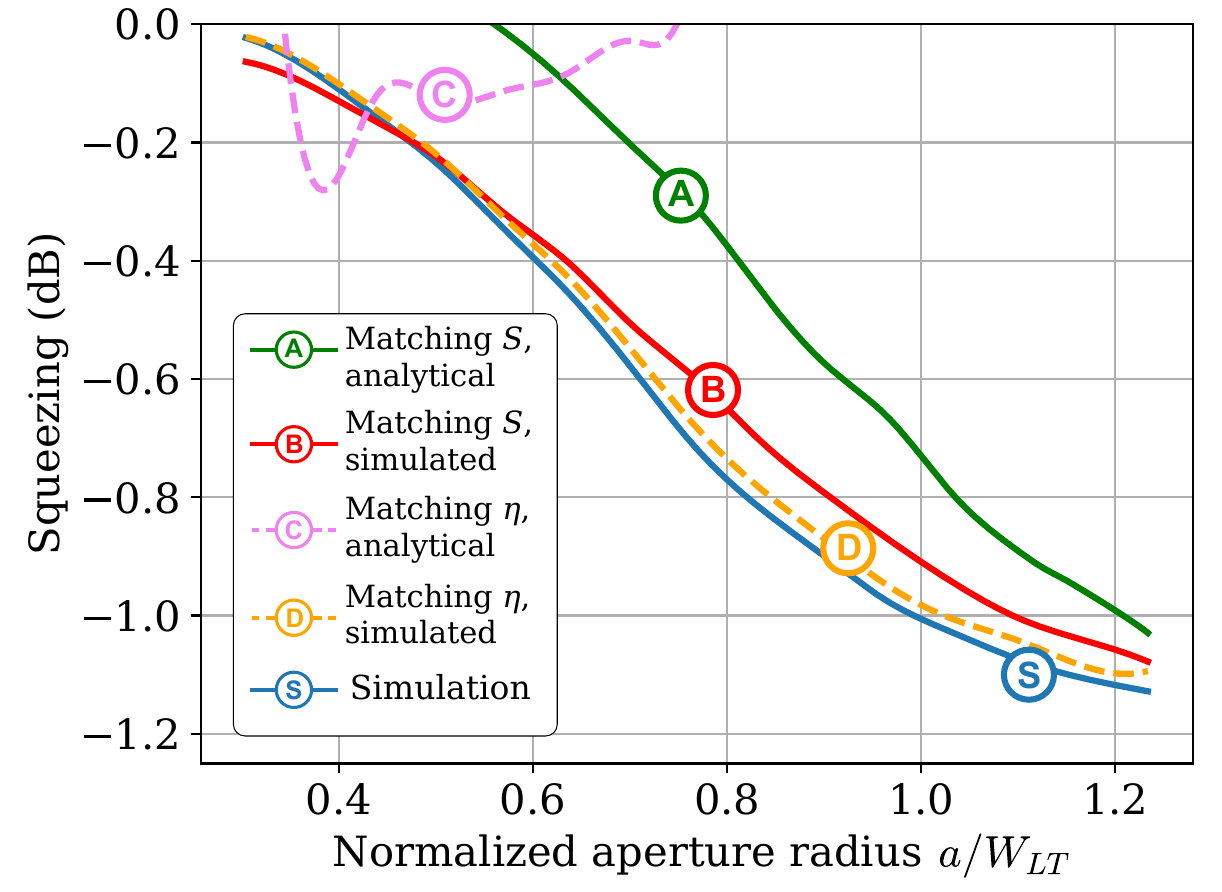}
        \caption{\label{Fig:app_Squeezing} 
        Comparison of squeezing parameter calculated by means of simulated PDT and PDTs obtained using the circular-beam approximation, based on different matching methods and either analytical or numerically simulated parameters.
        For the legend, see Fig.~\ref{Fig:app_Mandel+Binomial}.
        The channel parameters are the same as in Fig.~\ref{Fig:PDT_distrib}(a), with aperture radius $a$ ranging from $8$ to $33$~mm, with $\alpha_0=4$, $\eta_\text{c}=0.48$, and initial squeezing $-3$~dB.
        } 
    \end{figure}

Figure~\ref{Fig:app_Squeezing} shows the quadrature squeezing for an amplitude-squeezed coherent state.
The transmittance–moment matching method using simulated moments exhibits only a small discrepancy compared to the simulated data.
However, the same matching method with analytical moments fails for nearly all aperture radius values.
The reason for this behavior is very similar to the discrepancy observed for the Mandel parameter.
Since the second term in Eq.~(\ref{Eq:QuadratInOut}) becomes large for high values of $\alpha_0$, even small errors in $\left\langle  \Delta T^2  \right\rangle$ cause significant errors in the resulting squeezing.
A further contributing factor to this discrepancy is the significant error in the analytical expressions for $\langle \eta\rangle$ at large aperture radii.
When matching the first two moments of $S$, the result is again quantitatively correct but shows a larger discrepancy for analytical values compared to numerical simulations.

\section{Summary and conclusions}
\label{Sec:Conclusions}

Analytical PDT models play a crucial role in analyzing the transfer of nonclassical effects through free-space channels and in quantum communication protocols implemented in such environments.
It is also well known that the applicability of these models strongly depends on several factors---one of the most important being the aperture radius.
While heuristic models such as the beta PDT can yield acceptable results over a wide range of aperture sizes, physically motivated models based on the actual behavior of beams propagating through the atmosphere are of particular interest due to their stronger physical justification.

The elliptic-beam model is an example of a physically motivated approach.
While it provides a good fit to both experimental and numerical data, the parameter estimates obtained through such fitting can be significantly biased.
This is because, in a real atmospheric channel, the instantaneous beam shape is strongly non-Gaussian.
As a result, fitting its shape using the moments of the squared beam-spot radius may not yield the most accurate Gaussian approximation of the actual beam profile.
Another limitation of the elliptic-beam approximation is that it requires involved numerical computations and analytical approximations due to the elliptical shape of the beam.

The circular-beam approximation proposed in this paper addresses the two problems mentioned above.
First, we assume that the instantaneous beam profile is circular, which significantly simplifies the calculations.
Second, we propose an alternative parameter-matching method---designed to overcome the problem of model misspecification bias---by matching the model parameters to the first two moments of the transmittance rather than to the first two moments of the squared beam-spot radius.
While this approach introduces more numerical calculations into the model, it significantly extends the range of its applicability across different aperture radii.

To validate our model, we compare it with numerical data obtained using the phase–screen method.
The results demonstrate acceptable accuracy and at least reproduce the numerical trends quantitatively.
The method based on matching the moments of the squared beam-spot radius shows good agreement for aperture radii around 1.25 times the long-term beam radius---similar to the validity range of the elliptic-beam and beam-wandering models.
For other aperture sizes, it is more reasonable to use the transmittance–moment matching method.

We have also found that the results are highly sensitive to errors in the calculated moments.
Consequently, inaccuracies in the analytical expressions for the moments---either $\langle S \rangle$ and $\langle S^2 \rangle$ (for the squared beam-spot radius moment matching method), or $\langle \eta \rangle$ and $\langle \eta^2 \rangle$ (for the transmittance–moment matching method)---can lead to significant discrepancies in the resulting PDT.
For our analysis, we used the phase approximation of the Huygens–Kirchhoff method and found that for certain parameter ranges, its accuracy was insufficient.
In this paper we have considered only the case of weak turbulence, since strong turbulence is characterized by even larger inaccuracies in the parameter values, although the model itself may still remain valid.

To demonstrate the applicability of our method, we examined its ability to describe the transmission of nonclassical properties of quantum light---specifically, sub-Poissonian and sub-binomial photocounting statistics, as well as quadrature squeezing---through free-space channels.
In all cases the method based on matching the moments of the squared beam-spot radius yields quantitatively correct results, although the discrepancy with numerical data may be somewhat larger.
The transmittance–moment matching method, by contrast, has the potential to deliver highly accurate results.
By using numerically simulated moments, we showed that this approach agrees well with the numerical data, indicating that the circular-beam PDT accurately reproduces the full numerical PDT beyond just the first two transmittance moments.

However, a challenge arises for bright nonclassical states---for instance, amplitude-squeezed coherent states with high coherent amplitude.
In such cases, even small errors in the variances $\langle \Delta\eta^2 \rangle$ (for sub-Poissonian light) and $\langle \Delta T^2 \rangle$ (for quadrature squeezing) can lead to large deviations in the output nonclassicality certifiers.
This highlights the need for more accurate analytical techniques when calculating the transmittance moments if the method is to be applied reliably in such regimes.
  
\section*{Acknowledgment}

The authors thank D. Vasylyev for enlightening discussions.
I.P. and A.A.S. acknowledge support from the National Academy of Sciences of Ukraine through Project No. 0125U000031 and from Simons Foundation International SFI-PD-Ukraine-00014573, PI LB.
M.K. acknowledges support from the National Research Foundation of Ukraine under Project No. 2023.03/0165.

\section*{Data availability}

The data that support the findings of this article are openly available at Ref.~\cite{pechonkin2025Data}.

\appendix


\section{Matching moments of squared beam-spot radius}
\label{App:MatchS}

In this section we describe the method of matching the moments of the squared beam-spot radius. 
Similar to the transmittance-moment matching, this method is employed to determine the parameters of the log-normal distribution, $\mu$ and $\sigma^2$, introduced in Eq.~(\ref{Eq:lognorm}).
These parameters are expressed in terms of the first two moments, $\left\langle S \right\rangle$ and $\left\langle S^2 \right\rangle$, as
\begin{align}\label{Eq:mu}
	\mu=\ln{\left(\frac{\left\langle S \right\rangle^2}{\sqrt{\left\langle S^2 \right\rangle}}\right)},
\end{align}
\begin{align}\label{Eq:sigma}
	\sigma ^2=\ln{\left(\frac{\left\langle S^2 \right\rangle}{\left\langle S \right\rangle^2}\right)}.
\end{align}
While the first moment, $\left\langle S \right\rangle$, is given by Eq.~(\ref{Eq:S}), the second moment, $\left\langle S^2 \right\rangle$, requires a more involved analysis.

For deriving $\left\langle S^2 \right\rangle$, we start by squaring and averaging the instant beam-spot radius (\ref{Eq:simpleS}):
\begin{align}\label{Eq:S**2}
	\left\langle S^2 \right\rangle = 16\bigg( \int_{\mathds{R}^4} d^2\textbf{r}_1 d^2\textbf{r}_2 x_1^2 x_2^2 \Gamma_4(\textbf{r}_1, \textbf{r}_2;L)\\
	- 2 \left\langle x_0^2 \int_{\mathds{R}^2} d^2\textbf{r} x^2 I(\textbf{r};L) \right\rangle   +  \left\langle x_0^4 \right\rangle\bigg).\nonumber
\end{align}
The difficulty with this expression lies in the fact that the second and third terms involve sixth- and eighth-order field correlation functions, which pose challenges for analytical calculations.
To overcome this issue, we will apply reasonable approximations, similar to those used in the elliptic-beam approximation.

Let us analyze the second term in Eq.~\eqref{Eq:S**2}. It depends on the sixth-order field correlation function. A straightforward approach would be to assume that the field intensities at three points in the transverse plane are approximately distributed according to the Gaussian law. Under this assumption, the sixth-order correlation function can be expressed in terms of the second- and fourth-order correlation functions. However, our numerical simulations show that this approach does not provide satisfactory accuracy.   

\begin{figure}[ht!]
	\centering
	\includegraphics[width=1\linewidth]{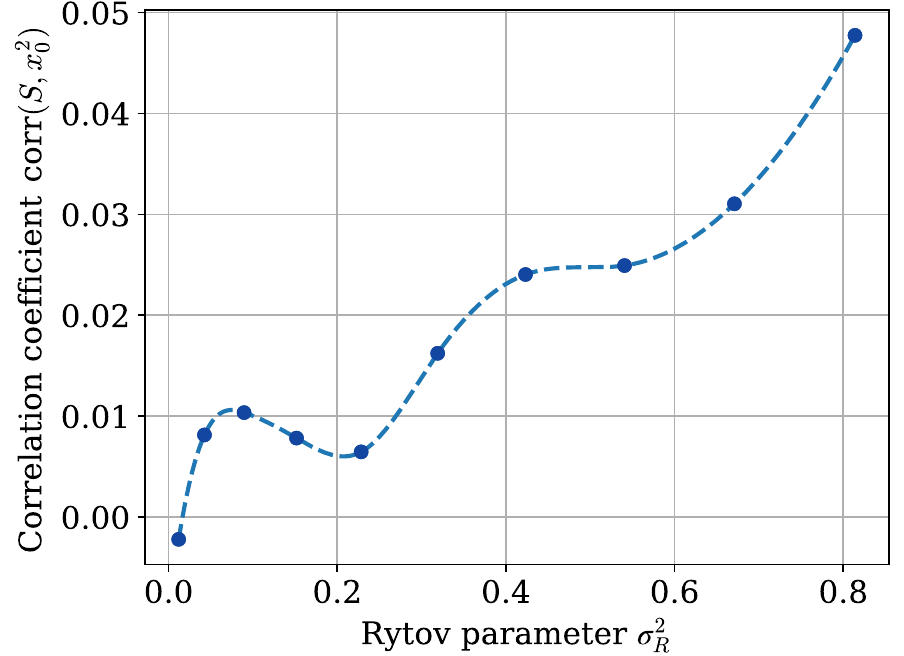}
	\caption{\label{Fig:CorrSx02} The Pearson correlation coefficient between $x_0^2$ and $S$, cf. Eq.~(\ref{Eq:Pearson}), as a function of the Rytov parameter.
		For the rest of the channel parameters, see Fig.~\ref{Fig:distrS}.}
\end{figure}

An alternative technique, used also in Ref.~\cite{vasylyev16}, consists in using Eq.~(\ref{Eq:simpleS}) to represent the second term in Eq.~(\ref{Eq:S**2}) as
\begin{align}\label{Eq:secondTerm}
	\left\langle x_0^2 \int_{\mathds{R}^2} d^2\textbf{r} x^2 I(\textbf{r};L) \right\rangle= 	\frac{1}{4}\left\langle x_0^2 S \right\rangle+\left\langle x_0^4 \right\rangle.
\end{align}
Our central assumption is that the correlation between $x_0^2$ and $S$ is weak.
This is confirmed by our numerical simulations, which evaluate the corresponding Pearson correlation coefficient between these variables,
\begin{align}\label{Eq:Pearson}
	\corr(S,x_0^2)=\frac{\left\langle Sx_0^2\right\rangle-\left\langle S\right\rangle\left\langle x_0^2\right\rangle}{\sqrt{\left(\left\langle S^2\right\rangle-\left\langle S\right\rangle^2\right)\left(\left\langle x_0^4\right\rangle-\left\langle x_0^2\right\rangle^2\right)}},
\end{align}
see Fig.~\ref{Fig:CorrSx02}.
The absolute value of the coefficient remains significantly below unity, thereby justifying the assumption of weak correlation.
This yields
\begin{align}\label{Eq:secondTermSeparation}
	\left\langle x_0^2 S \right\rangle\approx \left\langle x_0^2 \right\rangle\left\langle S \right\rangle.
\end{align} 
Finally, we take into account that the beam-centroid coordinate $x_0$ can be considered Gaussian distributed with high accuracy \cite{klen2023}.
This implies that
\begin{align}\label{Eq:xGauss}
	\left\langle x_0^4 \right\rangle\approx3\left\langle x_0^2 \right\rangle^2.
\end{align}
By combining all these assumptions in Eq.~(\ref{Eq:S**2}), we arrive at its approximate form
\begin{align}\label{Eq:S**2simplif}
	\left\langle S^2 \right\rangle \approx 16\bigg( \int_{\mathds{R}^4} d^2\textbf{r}_1 d^2\textbf{r}_2 x_1^2 x_2^2 \Gamma_4(\textbf{r}_1, \textbf{r}_2;L)\\
	- \frac{1}{2}\left\langle x_0^2 \right\rangle\left\langle S \right\rangle-3\left\langle x_0^2 \right\rangle^2     \bigg),\nonumber
\end{align}
which requires only the knowledge of the second- and fourth-order field correlation function.

\section{Analytical parameters from the phase approximation of the Huygens–Kirchhoff method}
\label{App:AnalyticalFormulas}

In this section we present the derivation of analytical expressions for the key parameters required in the circular-beam approximation.
A central quantity is the variance of the beam-centroid coordinate, $\sigma_{\text{bw}}^2$.
The moments $\left\langle S\right\rangle$ and $\left\langle S^2\right\rangle$ are needed for the parameter-matching technique based on the first two moments of $S$, as described in Appendix~\ref{App:MatchS}.
These parameters also provide the initial guess for the iterative procedure used to determine 
$\mu$ and $\sigma^2$ in the transmittance-moment matching method.
The latter additionally requires the transmittance moments $\left\langle \eta\right\rangle$ and $\left\langle \eta^2\right\rangle$.	
The core idea behind these derivations is to substitute the field correlation functions $\Gamma_2(\mathbf{r};L)$ and $\Gamma_4(\mathbf{r}_1,\mathbf{r}_2;L)$---see Eqs.~(\ref{Eq:Gamma2approx}) and (\ref{Eq:Gamma4approx})---into the corresponding definitions of these quantities.
The main challenge lies in finding suitable approximations for the resulting integrals.

To derive the analytical expression for $\langle S \rangle$, we substitute Eqs.~(\ref{Eq:Gamma2approx}) and (\ref{Eq:structureF}) into Eq.~(\ref{Eq:S}).
We start with the first term in Eq.~(\ref{Eq:S}), for which we use the dimensionless variables $\textbf{q}$ and $\textbf{q}^\prime$ via the substitution
	\begin{align}\label{Eq:substitutions1}
        \begin{cases}
		\textbf{r} = \textbf{q}W_0/\Omega\\
        \textbf{r}^\prime= \textbf{q}^\prime W_0/\Omega
        \end{cases}.
	\end{align}
As a common approximation, we replace $(\mathbf{q}'/\Omega)^{5/3}$ with $(\mathbf{q}'/\Omega)^2$, which is valid in the regime of weak turbulence.
The remaining steps involve evaluating Gaussian integrals. 
Upon carrying out the calculations, we obtain the expression
	\begin{align}\label{Eq:SFinal}
		\left\langle S \right\rangle= W_0^2\Omega^{-2}+2.93 & W_0^2 \sigma_R^2\Omega^{-7/6}\\
		&+0.24  W_0^2 	\sigma_R^4\Omega^{-1/3}.\nonumber
	\end{align}
 Here $\sigma_R^2$ is the Rytov parameter.

To evaluate $\sigma_{\text{bw}}^2$ [which also corresponds to the second term in Eq.~(\ref{Eq:S})] and $\langle S^2\rangle$ we take a different approach. 
We substitute Eqs.~(\ref{Eq:Gamma4approx}) and (\ref{Eq:structureF}) into Eqs.~(\ref{Eq:bwVariance}) and (\ref{Eq:S**2simplif}). 
The subsequent calculations reduce to evaluating two similar integrals: one for $\sigma_{\text{bw}}^2$ and another one for the first term in Eq.~(\ref{Eq:S**2simplif}).   
It is convenient to introduce the new variables $\textbf{R}$ and $\boldsymbol{\rho}$,
	\begin{align}\label{Eq:substitutions2}
        \begin{cases}
		(\textbf{r}_1+\textbf{r}_2)/2 = \textbf{R},\\
        \textbf{r}_1-\textbf{r}_2 = \boldsymbol{\rho},
        \end{cases}
	\end{align}
and make use of the identity involving the Dirac $\delta$ function,
	\begin{align}\label{Eq:Dirac}
		\int_{\mathds{R}^4} d^2\textbf{X} d^2\textbf{r}\, \textbf{X}^{2n} e^{-i\gamma\textbf{X}\cdot \textbf{r}}f\left(\textbf{r}\right)=\frac{(2\pi)^2}{\gamma^2} \left(-\frac{\Delta_\textbf{r}}{\gamma^2}\right)^n f(\textbf{r})\Bigg|_{\textbf{r}=0},
	\end{align}
where $\Delta_\textbf{r}$ is the transverse Laplace operator. 

One of the exponential terms containing the structure function [see Eq.~(\ref{Eq:Gamma4approx})], after substitution and rearrangement, has the argument
	\begin{align}\label{Eq:lastExp1}
		\rho_0^{-5/3} \int_0^1 d\xi & \Big[ \sum_{j=1,2}|\boldsymbol{\rho}\xi+(\textbf{r}_1^\prime+(-1)^j\textbf{r}_2^\prime)(1-\xi)|^{5/3}\\
        & - 2|\boldsymbol{\rho}\xi+\textbf{r}_1^\prime(1-\xi)|^{5/3}-2|\textbf{r}_2^\prime(1-\xi)|^{5/3}\Big]. \nonumber
	\end{align} 
In the weak–turbulence regime, this expression remains small.
Therefore, the terms involving $\boldsymbol{\rho}$ can be neglected and the expression simplifies to
    \begin{align}\label{Eq:structureFapprox}
		\rho_0^{-5/3} \int_0^1 d\xi & \Big[ \sum_{j=1,2}|(\textbf{r}_1^\prime+(-1)^j\textbf{r}_2^\prime)(1-\xi)|^{5/3}\\
        & - 2|\textbf{r}_1^\prime(1-\xi)|^{5/3}-2|\textbf{r}_2^\prime(1-\xi)|^{5/3}\Big]. \nonumber
	\end{align}	
We then expand the exponential with respect to $\rho_0^{-5/3}$ up to the second order. 
Where appropriate, Eq.~(\ref{Eq:Dirac}) is employed, after which the remaining integrals can be expressed in terms of Gamma functions and Gaussian integrals. As a result, we arrive at the expression for the second moment:
	\begin{align}\label{Eq:S2Final}
		\left\langle S^2 \right\rangle= &W_0^4\Omega^{-4}+6.48 W_0^4 \sigma_R^2\Omega^{-19/6}+9.40 W_0^4 \sigma_R^4\Omega^{-7/3}\nonumber\\
		&+2.60W_0^4 \sigma_R^6\Omega^{-3/2}-0.05W_0^4 \sigma_R^8\Omega^{-2/3}.
	\end{align} 
Finally, the closed-form expression for $\sigma_{\text{bw}}^2$ is summarized in Eq.~(\ref{Eq:bwFinal}).

\begin{figure}[ht!]
	\centering
	\includegraphics[width=1\linewidth]{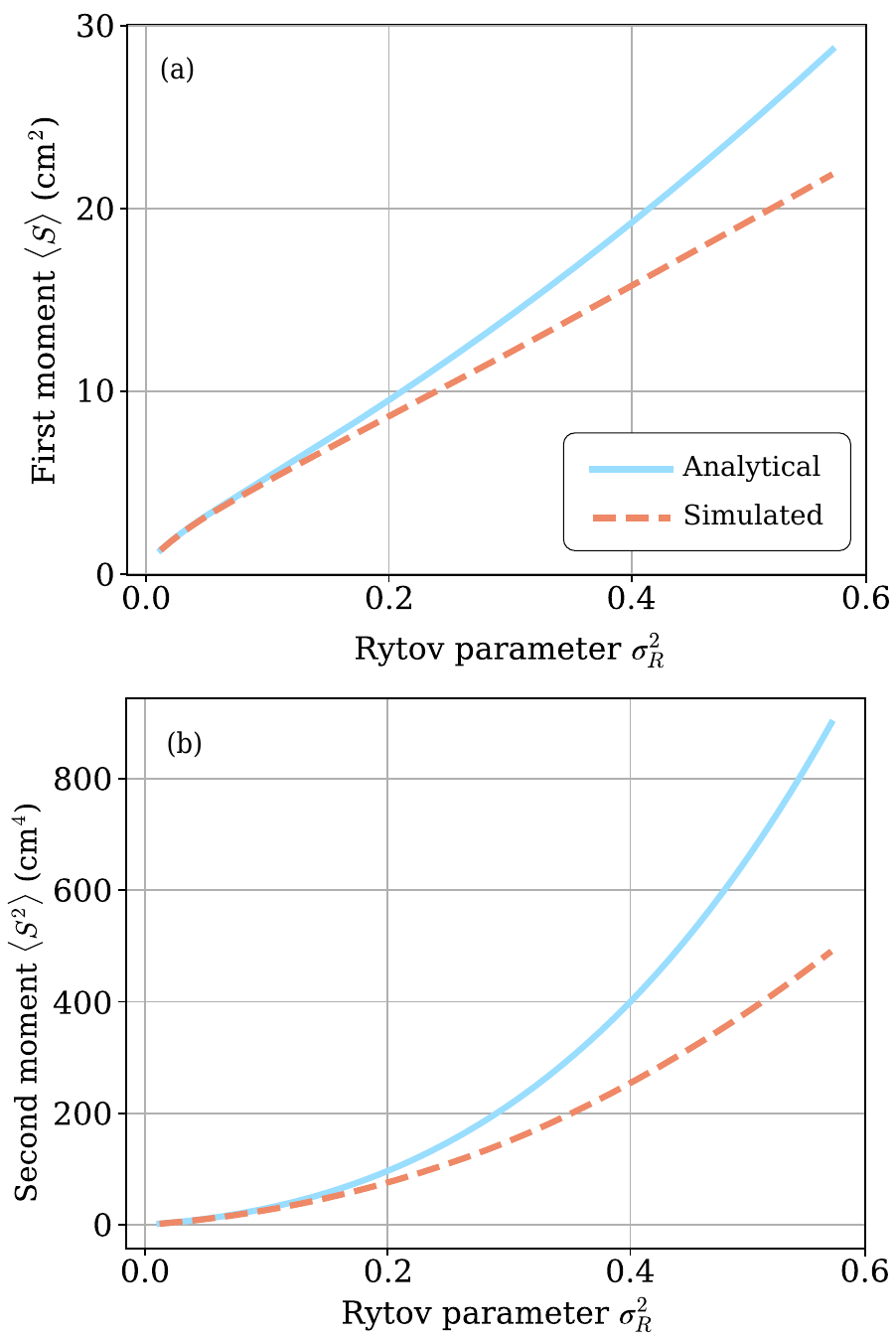}
	\caption{\label{Fig:comp_S} Comparison of analytical results obtained using the Huygens–Kirchhoff method (solid blue lines) with simulated data based on the sparse-spectrum model of the phase-screen method (dashed orange lines) for (a)~$\left\langle S \right\rangle$ and (b)~$\left\langle S^2 \right\rangle$.
	The channel parameters are the same as in Fig.~\ref{Fig:sim-analy}(d).}
\end{figure}

Similar to Sec.~\ref{Sec:Analytical}, to assess the validity of the analytical expressions for $\left\langle S \right\rangle$ and $\left\langle S^2 \right\rangle$, we compare them with numerical results obtained using the sparse-spectrum implementation of the phase-screen method.
The comparison is shown in Fig.~\ref{Fig:comp_S} for the first and second moments of the squared beam-spot radius.
Good agreement is observed only for small values of $\sigma_R^2$.

To evaluate the first moment of transmittance, $\left\langle \eta\right\rangle$, we substitute Eqs.~(\ref{Eq:Gamma2approx}) and (\ref{Eq:structureF}) into Eq.~(\ref{Eq:etaAvrg}) and apply the change of variables from Eq.~(\ref{Eq:substitutions1}). 
Next, we approximate $\left|\textbf{q}/\Omega \right|^{5/3}$ by expanding it with respect to $\left(\textbf{q}/\Omega \right)^{2}-1$:
	\begin{align}\label{Eq:appr}
		\left|\frac{\textbf{q}}{\Omega}\right|^{5/3}\approx 1+\frac{5}{6}\left(\left(\frac{\textbf{q}}{\Omega}\right)^{2}-1\right).
	\end{align}
After performing the integrations, we obtain
    \begin{align}\label{Eq:etaBeforeFinal}
		\left\langle\eta\right\rangle = & \exp\left(-0.13\sigma_R^2\Omega^{-5/6}\right)\\
        &\times\left[ 1-\exp\left( -\frac{a^2}{0.5 W_0^2\Omega^{-2}+0.66 W_0^2\sigma_R^2\Omega^{7/6}} \right)\right]. \nonumber
	\end{align}
Since the first exponential factor is close to one in the weak–turbulence regime, we neglect it to ensure that $\left\langle \eta \right\rangle \to 1$ as $a \to \infty$.

To evaluate $\langle\eta^2\rangle$, we substitute Eqs.~(\ref{Eq:Gamma4approx}) and (\ref{Eq:structureF}) into Eq.~(\ref{Eq:eta2Avrg}), then sequentially apply the variable substitutions from Eqs.~(\ref{Eq:substitutions2}) and (\ref{Eq:substitutions1}) for all integration variables, followed by the same approximation used in Eq.~(\ref{Eq:structureFapprox}).
The exponent involving the structure function then becomes
	\begin{align}\label{Eq:lastExp2}
		\frac{3}{8}\rho_0^{-5/3} & W_0^{5/3} \sum_{j=1,2}\Bigg[\bigg|\frac{\textbf{q}_1^\prime+(-1)^j\textbf{q}_2^\prime}{\Omega}\bigg|^{5/3}\\
        & - \bigg|\frac{\textbf{q}_2^\prime+(-1)^j\textbf{q}_3^\prime}{\Omega}\bigg|^{5/3} - \bigg|\frac{\textbf{q}_1^\prime+(-1)^j\textbf{q}_3^\prime}{\Omega}\bigg|^{5/3}\Bigg]. \nonumber
	\end{align} 
We approximate this expression using the same technique as discussed after Eq.~(\ref{Eq:substitutions1}):
	\begin{align}\label{Eq:lastExp3}
		\frac{3}{8} & \rho_0^{-5/3} W_0^{5/3} \sum_{j=1,2}\Bigg[\bigg(\frac{\textbf{q}_1^\prime+(-1)^j\textbf{q}_2^\prime}{\Omega}\bigg)^{2}\\
        & - \bigg(\frac{\textbf{q}_2^\prime+(-1)^j\textbf{q}_3^\prime}{\Omega}\bigg)^{2} - \bigg(\frac{\textbf{q}_1^\prime+(-1)^j\textbf{q}_3^\prime}{\Omega}\bigg)^{2}\Bigg]=\nonumber\\
        &=-\frac{3}{2}\rho_0^{-5/3} W_0^{5/3}\Omega^{-2} \textbf{q}_3^{\prime2}.\nonumber
	\end{align} 
The next step involves performing integrations that yield the modified Bessel function of the first kind and zero order, $\BesselM_0(z)$. 
One of the integrals can be expressed in terms of a function similar to the Rice cumulative distribution, $\mathcal{F}(r; \sigma, \nu)$,
    \begin{align}\label{Eq:Rice}
		\mathcal{F}(r; \sigma, \nu)=&\int_0^rdr^\prime \\
        &\times \frac{r^\prime}{\sigma^2}\exp\left(-\frac{r^{\prime2}+\nu^2}{2\sigma^2} \right) \BesselM_0\left(\frac{r^\prime \nu}{\sigma^2}\right). \nonumber
	\end{align} 
Following Ref.~\cite{Vasylyev2013}, we use an approximation of the Rice cumulative distribution.
Since the first-order approximation does not significantly improve the result, we adopt the simpler zero-order approximation:
    \begin{align}\label{Eq:RiceAppr}
		\mathcal{F}(r; \sigma, \nu)\approx 1-\exp\left(-\frac{r^2}{2\sigma^2}\right). 
	\end{align}
The remaining integrations are evaluated using standard Gaussian integral techniques.

\bibliography{biblio}

\end{document}